\documentclass[twocolumn]{emulateapj}
\usepackage{epsfig}
\usepackage{amsmath}

\newcommand{\Msolar}{M$_{\odot}$}
\newcommand{\Rsolar}{R$_{\odot}$}
\newcommand{\kms}{km s$^{-1}$}
\newcommand{\eoi}{$e/i$}

\newcommand{\tvm}{129}    %BM+BLM
\newcommand{\orbm}{85}    %All orbit solutions with PM>0 and RV>0

\newcommand{\magn}{$10.8 \leq V \leq 16.5$}         %Magnitude range
\begin{document}

\title{WIYN Open Cluster Study. XLVIII. The Hard-Binary Population of NGC 188}
\shorttitle{WOCS. The Hard-Binary Population of NGC 188}

\author{Aaron M. Geller\footnote{Visiting Astronomer, Kitt Peak National Observatory, National Optical Astronomy Observatory, which is operated by the Association of Universities for Research in Astronomy (AURA) under cooperative agreement with the National Science Foundation.}}
\affil{Center for Interdisciplinary Exploration and Research in Astrophysics (CIERA) and Dept. of Physics and Astronomy, Northwestern University, 2145 Sheridan Rd, Evanston, IL 60208, USA}
\affil{Department of Astronomy, University of Wisconsin - Madison, WI 53706 USA}
\email{a-geller@northwestern.edu}

\and
\author{Robert D. Mathieu$^{*}$}
\affil{Department of Astronomy, University of Wisconsin - Madison, WI 53706 USA}
\email{mathieu@astro.wisc.edu}

\shortauthors{Geller \& Mathieu}

\begin{abstract}

We present an in-depth study of the hard-binary population of the old (7 Gyr) open cluster 
NGC 188.  Utilizing \orbm~spectroscopic binary orbits out of a complete sample of \tvm~detected binary
members, we study the cluster binary frequency and the distributions of binary orbital elements amongst the 
main-sequence, giant and blue straggler populations.  
The results are derived from our ongoing radial-velocity survey
of the cluster, which spans in magnitude from the brightest stars in the cluster 
to $V=16.5$ (about 1.1 - 0.9 \Msolar),
and extends to a projected radius of 17 pc ($\sim$13 core radii). Our detectable binaries have 
periods ranging from a few days to of order 10$^4$ days, and thus are hard binaries 
that dynamically power the cluster.  
The main-sequence solar-type hard binaries in NGC 188 are nearly indistinguishable from similar
binaries in the Galactic field.
We observe a global solar-type main-sequence hard-binary frequency in NGC 188 of 23~$\pm$~2~\%, 
which when corrected for incompleteness results in a frequency of 29~$\pm$~3~\% for 
binaries with periods less than 10$^4$ days.  
For main-sequence hard binaries in the cluster we observe a log-period distribution that rises towards our detection limit, 
a roughly Gaussian eccentricity distribution centered on $e = 0.35$
(for binaries with periods longer than the circularization period), and a secondary-mass distribution 
that rises towards lower-mass companions.
Importantly, the NGC 188 blue straggler binaries show significantly different characteristics than the
solar-type main-sequence binaries in NGC 188.
We observe a blue straggler hard-binary frequency of 76~$\pm$~19~\%, three times that of the
main sequence. The excess of this binary frequency
over the normal main-sequence binary frequency is valid at
the $>$99\% confidence level.
Furthermore, the blue straggler binary eccentricity - log period distribution is 
distinct from that of the main-sequence at the 99\% confidence level, with the majority of the blue straggler binaries
having periods of order 1000 days and lower eccentricities.   
The secondary-mass distribution for these long-period blue straggler binaries 
is narrow and peaked with a mean value of about 0.5 \Msolar.
Predictions for mass-transfer products are most closely consistent with the binary properties of these
NGC 188 blue stragglers, which comprise two-thirds of the blue straggler population.
Additionally we compare the NGC 188 binaries to those evolved within the sophisticated \citet{hur05} 
$N$-body open cluster simulation.  The main-sequence hard-binary population predicted by the simulation 
is significantly different from the main-sequence hard-binary population observed in NGC 188, in frequency
and distributions of period and eccentricity.  Many of these differences result from the adopted 
initial binary population, while others reflect on the physics used in the simulation (e.g., tidal circularization).
Additional simulations with initial conditions
that are better motivated by observations are necessary to properly investigate the dynamical evolution
of a rich binary population in open clusters like NGC 188.
\end{abstract}


\section{Introduction}

Binaries influence the evolution of open clusters and are catalysts for the formation of 
anomalous stars like blue stragglers (BSs).
Stellar dynamicists have long recognized that encounters 
involving binary stars can supply the inevitable energy flow out of the cores of star clusters.  
These complex dynamical dances also lead to stellar exchanges, binary orbit evolution, close stellar 
passages, mass transfer, mergers, and stellar collisions, all of which lead to an array of 
new stellar evolution paths and products.  Observations of the old \citep[7 Gyr;][and see also \citealt{mei09}]{sar99} open cluster NGC 188 
reveal a rich binary population and a wide variety of such interaction-product candidates, 
including BSs, sub-subgiants and X-ray sources 
\citep[e.g.,][]{bel98,gon05,gel08b,gel09}.  This abundance of anomalous stars may be a
direct result of the dynamical evolution of the single and binary stars as they interacted
over the lifetime of the cluster.

The field binaries are an important comparison population to open cluster binaries, as
we would not expect field binaries to undergo dynamical encounters due to the low stellar 
density of the Galactic field.
Therefore comparisons between open cluster and field binary populations may provide insights 
into the impact of stellar dynamics on shaping a binary population and creating exotic stars.
There have been a number of studies of the field dwarf binaries in the past \citep[notably, ][]{abt76,duq91}.
Currently the most comprehensive and complete survey of field dwarf binaries is that of \citet[][hereafter R10]{rag10}.
They analyze the multiplicity of a complete sample of 454 $\sim$F6 - K3 dwarf and subdwarf stars
within 25 pc of the Sun, finding a frequency of multiple systems of 46\% (34\% $\pm$ 2\% strictly binaries).
The orbital period distribution is fit by a log-normal curve centered on log($P$ [days] ) = 5.03.
For binaries with periods longer than the circularization period, R10 find that the orbital eccentricity distribution is 
roughly flat out to $e \sim 0.6$.
The mass-ratio distribution for binaries is found to be roughly uniform between 0.2 and 0.95, 
with an additional peak at a mass ratio of unity (commonly termed ``twins'').

\citet[][hereafter Paper 2]{gel09} present \orbm~solar-type binary orbits in NGC 188 derived from 
our ongoing radial-velocity (RV) survey of the cluster. These binaries are of similar spectral type to those 
studied by R10, and thus provide a comparison between field binaries and those that have evolved 
within the dynamical environment of an open cluster.
This work is an integral part of the WIYN\footnote{\footnotesize
The WIYN Observatory is a joint facility of the University of Wisconsin - Madison, Indiana University, Yale
University, and the National Optical Astronomy Observatories.} Open Cluster Study \citep[WOCS;][]{mat00}.
Our observations cover a complete sample of stars within
a magnitude range of \magn~(1.1 - 0.9 \Msolar) and extending to 30 arcmin (17 pc in projection or roughly 13 core 
radii)\footnote{We adopt a core radius of 1.3 pc \citep{bon05} at a distance of 1.9 kpc, which corresponds 
to 2.35 arcminutes on the sky.} in radius, essentially out to the tidal radius of the cluster 
\citep[21 $\pm$ 4 pc;][]{bon05}. 
Our magnitude limits include solar-type main-sequence (MS) stars, subgiants, giants, and BSs.
For most binaries, our observations span $\sim$11 years, while for some binaries our time baseline covers
up to 35 years.  Our detectable binaries have periods ranging from a few days to of order 10$^4$ days.
A full description of our stellar sample, observations, data reduction routine, and precision,
as well as a summary table listing relevant information on each star in our sample, can be found 
in \citet{gel08b} (hereafter, Paper 1).
The binary orbital parameters, including estimates for the primary masses, can be 
found in Paper 2. 

In this third paper in the series, we study the characteristics of the NGC 188 binary population.
We first perform a detailed analysis of our observational completeness (Section~\ref{incomp}), 
and use the results from this section to study the global hard-binary frequency (Section~\ref{freq}).
In Section~\ref{SelogP} we discuss the eccentricity - log period ($e - \log(P)$) diagram.  
We then analyze the hard-binary distributions in eccentricity 
(Section~\ref{Sefreq}), period (Section~\ref{SPfreq}),   mass ratio and secondary mass 
(for MS binaries in Section~\ref{Sqm2freq} and BS binaries in Section~\ref{Sqm2freqBS}).  
Throughout the paper, we compare the MS, giant and BS binary populations.
Additionally, we compare the NGC 188 binary population to similar binaries in the Galactic field.
In Section~\ref{Hcomp}, we also compare these results to the $N$-body open cluster simulation of \citet{hur05},
and we build on the work of \citet{mat09} and \citet{gel11} to discuss the 
possible origins of the NGC 188 BS population. Finally, in Section~\ref{summary} we provide a brief summary and conclusions.

\section{Completeness in Binary Detection and Orbital Elements} \label{incomp}

There are 630 stars with non-zero proper-motion membership probabilities from \citet{pla03} within the magnitude and 
spatial domain that define our complete stellar sample.  
Updating the results of Paper 1, we have obtained $\geq$3 RV measurements for 623 of these stars\footnote{We require at least three 
observations before we categorize a given star as either single or binary.
We use the term ``single'' to identify stars with no significant RV variation;
a star is termed single if the standard deviation of its RV measurements is less than four times our precision.
Certainly some of these stars are also binaries, although generally with longer periods and/or lower total mass than the binaries
identified in this study.}.  
Thus, we have derived RV membership probabilities and characterized the RV variability of 99\% of this complete sample.  
Two of the seven stars that don't have three observations are too blue to derive reliable RVs with our current observing setup 
and two are rotating too rapidly (both are W UMa's).
The stars with nonzero proper-motion and RV memberships define the stellar sample used in the subsequent analysis, 
totaling 487 primary stars.
Here, we characterize our completeness in binary detection and orbital solutions within this sample.

We use a Monte Carlo approach to produce simulated observations of 
a set of simulated binaries with characteristics of the Galactic field binary population of R10.  
Specifically we choose the binary periods from a log-normal distribution, centered on log($P$ [days] ) = 5.03 with 
$\sigma_{log P} = 2.28$.  
We fit a wide Gaussian to the observed eccentricity distribution in R10 that is centered on $e = 0.39$ with $\sigma_e = 0.31$
($\chi^2_{red} = 0.78$) and choose the eccentricities from this distribution.
(A flat eccentricity distribution fit to the R10 data results in $\chi^2_{red} = 2.28$.)
We use 1 \Msolar~MS primary stars and choose mass ratios ($q$) from a flat distribution such that
the secondary stars have masses between 0.08~-~1 \Msolar.  
We do not include the peak in the observed field distribution at mass ratios of unity as we detect such binaries
as double-lined systems in our spectra and identify them as binaries directly.  Once detected, we observe double-lined systems 
on a more frequent basis than the majority of the binaries in our survey (see Paper 1).  
The orbital inclinations and phases of the binaries are chosen randomly.  
To each of the simulated observations of a given simulated binary, we add a random error generated from a Gaussian 
distribution centered on zero and with a standard deviation equal to our single-measurement precision of 0.4 \kms.  

As defined in Paper 1, an observed star is considered to be in a binary if the standard 
deviation of its RV measurements is greater than four times our precision of 0.4 \kms~(or \eoi~$>$~4).  
In our survey, we initially followed \citet{mat83} in requiring three observations over the course of at least one year
before attempting to categorize a star as being either single or binary.  Here, we use our Monte Carlo
method to re-evaluate this procedure.  We determine our binary detection completeness by analyzing simulated observations
of the simulated binary population covering time intervals defined by our actual observations of NGC 188.

\begin{figure*}[!t]
\plotone{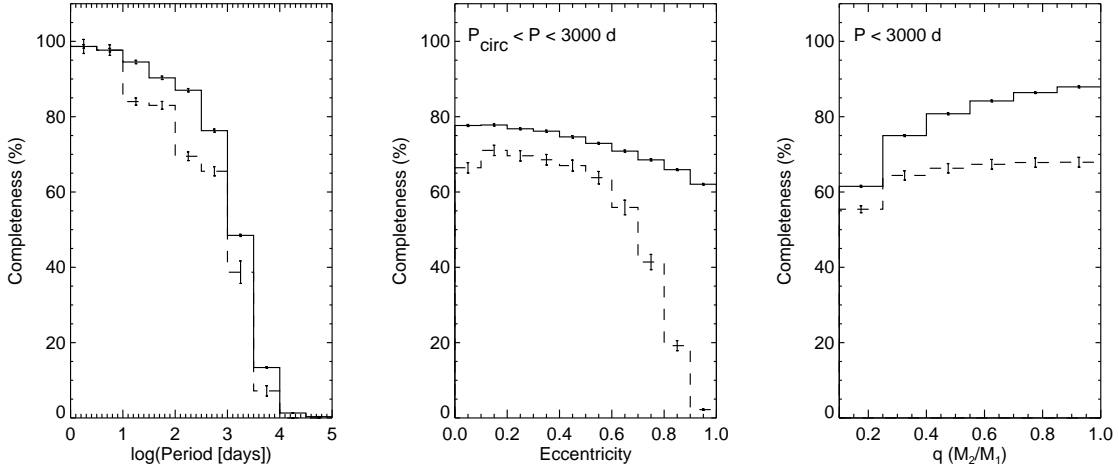}
\caption{\footnotesize  Observational completeness curves for NGC 188 binaries resulting from our Monte Carlo analysis.   
We plot our completeness as a function of period (left), eccentricity ($P_{circ} < P \leq 3000$ d; center) and mass ratio (P$\leq 3000$ d; 
right). In all plots we show our completeness in binary detection with the solid line and in orbital 
solutions with the dashed line.  
%These completeness curves are the result of our Monte Carlo analysis in which we
%``observe'' a simulated population of binaries with orbital parameters distributed according to the R10 field 
%population.
}
\epsscale{1.0}
\label{incfig}
\end{figure*}

Specifically, we have observed the stars in our NGC 188 WIYN sample over the course of 
about fourteen years.  The distribution of the time separation between our first 
two observations is bimodal with two equally strong peaks, one at about two months and another at about nine months.
The distribution of the time separation between our first and third observations is 
also bimodal, with a strong peak at about eight months and a smaller peak at about 2.5 years.  
These distributions reflect the approximate time between individual observing runs and the number of stars that 
we can observe during each observing run (see Paper 1 for a full description). We use these distributions
to define time intervals of simulated observations in our Monte Carlo analysis in order to determine our completeness in binary detections.

This analysis shows that 
with our first two observations we should detect 
78\%
of the binaries with periods less than 10$^3$ days.  If we include a third observation, this value increases significantly to 
88\%.
If we include a fourth observation, our completeness only increases to 
90\%.
Thus to optimize our efficiency and completeness in 
detected binaries, we require three observations spanning roughly one year before classifying a star as either single or 
binary, confirming the results of \citet{mat83}. 
We plot our binary detection completeness as functions of binary orbital elements in the solid lines in 
Figure~\ref{incfig}, and we use this analysis to correct our binary frequencies in Section~\ref{freq}.

To estimate our incompleteness in binary orbital solutions, we simulate observations of the simulated binaries 
on the actual observation dates of each of the NGC 188 binaries, respectively.
For each combination of a simulated binary with a set of true observation dates, we attempt to derive an 
orbital solution using an automated version of the same code used in Paper 2.  This code iteratively 
minimizes the residuals between the RV data and the orbital fit to return the best fitting orbital 
solution.  An acceptable orbital solution has RVs covering at least one period, errors on 
$P$, $e$ and the orbital amplitude $K$ of less than 30\% of the derived values, respectively, an RMS
residual velocity of less than 1 \kms, and a range in RVs covering at least 75\% of the orbital amplitude
(mainly applicable for highly eccentric binaries).
The resulting orbits define our completeness in orbital solutions, and each binary in NGC 188 produces 
a unique completeness curve in period, eccentricity, etc.
The completeness resulting from all combinations of actual observing dates and simulated binaries is plotted in the dashed 
lines in Figure~\ref{incfig}.
These histograms show the fraction of all binaries (within the specified period limits) for which we expect to derive orbital solutions, and 
hence obtain the orbital parameters.
We use this incompleteness analysis to correct the observed distributions in period, eccentricity, secondary mass and mass ratio.

Our completeness in detected binaries and orbital solutions decreases dramatically with increasing period
(see the left panel of Figure~\ref{incfig}).  
Longer-period binaries generally have a decreased amplitude of RV variations and 
require a longer time baseline of observations in order to detect and fully characterize the orbit.
We detect 
88\%
of binaries with periods less than 1000 days,
78\% of binaries with periods less than 3000 days, and
63\% 
of binaries with periods less than 10$^4$ days. 
Only a negligible fraction of binaries with periods greater than 10$^4$ days is detectable with our 
current data.   We therefore use the completeness in detected binaries with $P < 10^4$ days in our 
analysis of the binary frequency in Section~\ref{freq}.
Only two of our MS binaries with orbital solutions have periods greater than 3000 days. 
Therefore for the following analysis of the distributions of orbital parameters, we choose to limit 
our investigation to binaries with $P < 3000$ days.

We show our completeness as a function of eccentricity for binaries with $P_{circ} < P < 3000$ days
\citep[$P_{circ}$~=~15.0 days,][]{mat04} in the middle panel of Figure~\ref{incfig}.  
The completeness in detected binaries is much less sensitive to orbital eccentricity than it is to period.
However, our completeness in orbital solutions drops off rapidly at the highest eccentricities.
High eccentricity binaries require well-placed observations to define the orbital amplitude and eccentricity,
greatly increasing the difficulty in deriving reliable orbital solutions.

In the right panel of Figure~\ref{incfig}, we plot our completeness in the mass ratio ($q = M_2/M_1$) 
for binaries with $P<3000$ days.
Because we use a single primary mass of 1 \Msolar~in this Monte Carlo analysis, the completeness in 
secondary mass is identical to the completeness in mass ratio shown here.
Therefore, this curve also shows our decreasing completeness with decreasing secondary mass, as expected from
spectroscopic surveys of this type.  

The completeness curves in Figure~\ref{incfig} are not independent.  
Our detectability is most sensitive to period, since shorter-period binaries generally have higher-amplitude orbits. 
Hence this first bin in period shows the highest completeness, but the completeness is still less than 100\% due to 
short-period binaries with very high eccentricity and/or very low-mass secondaries.
Likewise no bin in eccentricity or mass ratio shows 100\% completeness due in part to long-period binaries found at 
all mass ratios and eccentricities.

We use this analysis to correct our observed MS and giant binary frequencies and distributions of orbital parameters for incompleteness.  
However we do not apply a completeness correction to the BS sample since we have generally observed the BSs more frequently
than the majority of the ``normal'' cluster stars, and
%(b) the true primary masses of the blue st
%BSs are presumably more massive than 1~\Msolar~(the turnoff massis $\sim$1.1~\Msolar), and 
the distributions of orbital parameters for the BSs may be significantly different than 
the field dwarfs and subdwarfs surveyed by R10.  There is only one detected binary BS for which we do not have 
an orbital solution (WOCS ID 8104).\footnote{\footnotesize  There is also one proper-motion member in the BS region of the color-magnitude 
diagram for which we have been unable to derive a reliable RV membership due to rapid rotation and RV variability (WOCS ID 4230).
The mean RV is outside of the cluster distribution, but we cannot be certain about membership without an 
orbital solution (and therefore a center-of-mass RV). We do not include this system in the following analysis due 
to it's uncertain membership; although we mention it here because it is an X-ray source 
\citep[S27,][]{gon05} and therefore warrants further observations.}
The orbital period for this binary is longer than the baseline of our observations.  This system is accounted for 
in our analysis of the BS binary frequency, but not in the distributions of BS orbital parameters.

Finally we note that there are only five non-RV-variable BSs in our sample.
If we coarsely estimate their primary masses by 
their proximity to evolutionary tracks \citep{mar08}, and use these primary masses along with the true 
observations of these BSs in our Monte Carlo analysis (simplistically applying the same distributions of orbital parameters as 
observed for field dwarfs), we find only a 24\% chance of having an undetected binary with 
a period $<10^4$ days in this sample.  Moreover we would only expect to have missed at most one BS binary in our data with $P<10^4$ days. 
Therefore any incompleteness correction must necessarily be minimal.

\section{Hard-Binary Frequency} \label{freq} 

A hard binary is defined as having an internal energy that is much greater than the energy of the relative 
motion, with respect to the binary, of a single star moving at the velocity dispersion of the cluster \citep{heg75}.
Using the method of \citet{gel10} to correct for our measurement precision and undetected binaries, 
we derive a one-dimensional velocity dispersion in NGC 188 of 0.41~$\substack{+0.09 \\ -0.10}$~\kms~for MS 
single members of NGC 188.  (This value is consistent with that found in Paper 1.)
Therefore, for solar-type stars in NGC 188, binaries with periods of less than $\sim$10$^6$ days 
are hard binaries.  As shown in Section~\ref{incomp}, nearly all of our detectable 
binaries have periods less than 10$^4$ days. Therefore we detect only hard binaries, and
in the following, we will use the term ``hard binary'' to refer specifically to the detectable hard binaries in NGC 188.
These hard binaries are catalysts for the production of anomalous stars like BSs and X-ray sources 
\citep[e.g.,][]{bai95,hur05,iva06}, 
and as such the hard-binary frequency may directly influence the number of such anomalous stars produced within 
an open cluster.
In this section, we compare the hard-binary frequencies of the NGC 188 
MS, giant and BS populations.  

We find 375 MS members including 90 MS binary members\footnote{Here, as in Paper 1, we use our 
binary member (BM) and binary likely member (BLM) classes to define our binary member sample.}. This observed 
MS hard-binary frequency is slightly inflated due to the faint magnitude limit defining our stellar sample.  
The combined light of two MS stars in a binary system can be up to 0.75 magnitudes 
brighter than the primary star alone.  Thus, the faint magnitude limit of our sample includes binary systems  
with primary stars that would lie below our magnitude limit.  Here we discard 
the four binaries that, after correcting for secondary light, have primaries below the faint limit.
We find an observed $V$ magnitude corrected MS hard-binary frequency of 23~$\pm$~2~\% (86/371).  
As shown in Section~\ref{incomp}, we detect 
63\%
of the MS binaries with periods of less than 10$^4$ days. 
The Monte Carlo analysis assumes that these binaries are detectable from the first three 
observations alone.  However continued observations of some stars in our sample have allowed us to 
detect certain (generally long-period) binaries after further observations, that did not show 
significant RV variability initially and therefore were not detected as binaries after only the first three observations.
Therefore, in order to correct 
this observed MS binary frequency for our incompleteness, we consider only the 
67
binaries that show significant RV variability (i.e., $e/i \geq 4$, as in Paper 1) in their first three observations.
When this sample is corrected for incompleteness,
we find that the true MS binary frequency among binaries with periods of less than 10$^4$ days is 
29~$\pm$~3~\%.

\citet{sol10} estimate a MS core binary frequency for NGC 188 of $58.2 \pm 13.5$ \% (over all mass ratios) through 
analysis of the CMD.  
To compare with this result, we first limit our sample to cover a similar spatial extent as that of \citet{sol10}.
Specifically, we select all stars within 17 arcminutes from the cluster center.  
(We note that \citet{sol10} included lower-mass primaries than are included in our RV survey, but we do not attempt 
to correct for this difference in primary mass range.)  Within our limited sample, we find a solar-type MS binary frequency of $30 \pm 4$ \% 
for binaries with periods $< 10^4$ days. If the NGC 188 solar-type MS binary period distribution is consistent with that of similar binaries in 
the Galactic field, then we can use the R10 field binary period distribution to extend this NGC 188 binary 
frequency out to the hard-soft boundary of the cluster, of $\sim 10^6$ days.  This analysis yields a MS binary frequency 
of $62 \pm 8$\%, within the given spatial domain and out to the cluster hard-soft boundary, in excellent agreement with the 
binary frequency derived by \citet{sol10}.

The NGC 188 hard-binary frequency also agrees with
values derived for other open clusters of a wide range of ages.  
\citet{mat90} find M67 (4 Gyr) to have a 
frequency of binaries with periods less than 10$^3$ days of 9\% to 15\%.  
Matching this period range, we derive a binary frequency of 22~$\pm$~3~\% out to a period of 10$^3$~days in NGC 188, 
somewhat higher than M67, but only at the margin of significance.
The similar WOCS study of the
intermediate-aged (2.5 Gyr) open cluster NGC 6819 \citep{hol09} report an observed binary frequency of 
17\%, all of which have periods less than 10$^4$ days.  We note that this observed binary frequency in NGC 6819
has not yet been corrected for incompleteness.  
The WOCS study of the young (150 Myr) open cluster M35 finds an incompleteness-corrected MS binary frequency of 24~$\pm$~3\%
for binaries with periods $<10^4$ days \citep{gel10}.
Additionally, \citet{mer08} find a spectroscopic binary frequency for stars of FGK spectral types 
of 20\% in Blanco 1 ($\sim$100 Myr).  Their observations have a similar time baseline to our NGC 188 
survey though with fewer epochs, suggesting a slightly lower detection frequency as found here.
Remarkably, these clusters span nearly 7 Gyr in age without a dramatic change in binary frequency.

\begin{figure}[!t]
\plotone{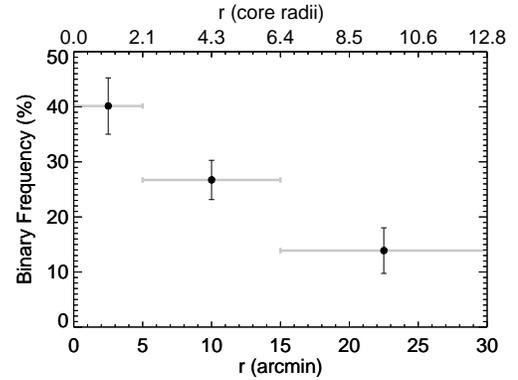}
\caption{\footnotesize  Incompleteness corrected hard-binary frequency as a function of projected radius, including 
the main-sequence and giant populations.  We remove faint main-sequence binaries with primary 
stars that would likely fall below our observational limit of $V=16.5$.
The vertical error bars show the one sigma Poisson errors on the 
binary frequencies.  The horizontal bars show the bin sizes of 5, 10 and 15 arcminutes, respectively.  
For main-sequence binaries we use our standard incompleteness correction, and for the giant binaries we apply a
minimum period of 11.5 days in our incompleteness correction.
We observe a central concentration of binaries as compared to the single stars in the cluster.}
\label{cbf}
\end{figure}

Out of the 454 primary stars in the R10 sample, there are 87 companions with periods less than 10$^4$ days, 
yielding a field binary frequency within this period range of 19~$\pm$~2~\%.
The NGC 188 hard-binary frequency out to the same 
period cutoff is marginally higher than that of the field.
This higher binary frequency in NGC 188 
may be a dynamical signature, as $N$-body models predict that single stars will be preferentially lost from a cluster 
through evaporation as compared to the generally higher total mass binaries \citep{hur05}.  
Thus NGC 188 may have started with a lower solar-type binary frequency, perhaps very similar to that of the field,
that increased throughout the cluster lifetime.
Furthermore, the hard-binary frequency in the halo of the cluster (see Figure~\ref{cbf}), where 
the dynamical relaxation time is long and stellar densities are low, is in good agreement
with that of the field.

We find 70 giants, 19 of which are in binaries, resulting in an observed giant 
hard-binary frequency of 27~$\pm$~6~\%.  (See Paper 1 for our definition of the giant region in the color-magnitude diagram, CMD.)
14 giant binaries would be detected after the first 3 observations alone, and are therefore applicable 
to our incompleteness correction.
As a star evolves along the giant branch, its increasing physical radius will preclude
more and more companions at close separations and short periods.  
\citet{mer07} find a minimum period of 41.5 days in their sample of red giant binaries in open clusters.
Using this as our minimum period in the incompleteness analysis, we find a giant binary frequency of 
37~$\pm$~10~\% for binaries with $P<10^4$ days.
The minimum period for the giants in our NGC 188 sample is $\sim$11.5 days.  Using this as the minimum period results in 
a binary frequency of 34~$\pm$~9~\% for binaries with $P<10^4$ days.
The giant hard-binary frequency derived using either minimum period is consistent with that of the MS within the 
same period range, and with too large an error to see evolutionary differences in the frequency.

In Figure~\ref{cbf} we plot the incompleteness corrected hard-binary frequency for 
the MS and giant populations as a function of projected radius from the cluster center.
Again, we remove the four faint MS binaries with primaries that likely fall below our $V$ magnitude limit.
We exclude the BSs because their spatial distribution may be tied to their 
formation mechanism(s) rather than dynamical (energy equipartition) effects.
We divide our data into three bins of increasing 
size, covering 5, 10 and 15 arcminutes, respectively, and denoted by the horizontal bars associated with each data
point. As shown in Figure~\ref{cbf} the binary frequency clearly decreases with radius from the cluster center.  
The binary frequency in the core is 
significantly higher than that of the halo; the first and last bins in Figure~\ref{cbf} are distinct with $>$99\% confidence.
This increase in binary frequency reflects the central concentration of the solar-type binaries with respect 
to the solar-type single stars in NGC 188 (as we also show in Paper 1).

In marked contrast to the MS and giant populations, \citet{mat09} find 16 of the 21 BSs to have binary companions.
(In Paper 1 we show the region on the CMD from which we have defined our BS population.)
These 16 BSs are secure binaries as 
we have derived orbital solutions for all but one; the remaining, 8104, is unquestionably a binary, 
with a longer period than our time baseline of observations for that star.  Thus, we find 
a BS hard-binary frequency of 76~$\pm$~19~\%, more than three times and distinct with $>$99\% confidence from the observed MS hard-binary 
frequency \citep{mat09}.  Furthermore, the five BSs that appear to be single may prove to be in longer-period 
binaries that currently lie outside of our spectroscopic detection limits.

As discussed in \citet{mat09}, the large BS binary frequency is not sensitive to our selection criteria.
For example, if we take only bright BSs ($V < 15$), the binary frequency is $76 \pm 21$~\% (13/17). 
Alternatively if we include only those BSs with proper-motion membership probabilities $\geq$90\% the binary 
frequency is $81 \pm 23$\% (13/16).  

We also note that there are two stars that lie above the MS turnoff and just fainter than our cutoff for BS selection (IDs 5101 and 8899).
Our selection criteria excludes any stars that could be members of normal MS binaries.
Thus, normal stars found just fainter than our BS region of the CMD are expected to be double-lined (SB2) binaries.  
To date we have not detected the flux from any companions in our spectra for either of these two stars, and their RVs 
do not vary above our threshold for binary detection.  Therefore both of these stars are considered single cluster 
members (SM, see Paper 1), and perhaps should be included in the BS sample.  If we include these two stars in the sample the 
BS binary frequency is $70 \pm 17$~\% (16/23), still significantly higher than the observed MS binary frequency.

In Paper 1 we find that the BSs appear to occupy a bimodal spatial distribution,
with a centrally concentrated population of 14 BSs and a halo population of 7 BSs. 
We compare the binary frequencies of the core and halo BS populations, divided at 10 arcmin from the cluster 
center, and find the core BSs to have a binary frequency of 79~$\pm$~24~\% (11/14) and the halo group to have a 
binary frequency of 71~$\pm$~32~\% (5/7).
The binary frequencies of these two subsets of NGC 188 BSs agree to within their large uncertainties.

\section{The \textit{e} - log(\textit{P}) Diagram} \label{SelogP} 

\begin{figure}[!t]
%\plotone{f3.eps}
\plotone{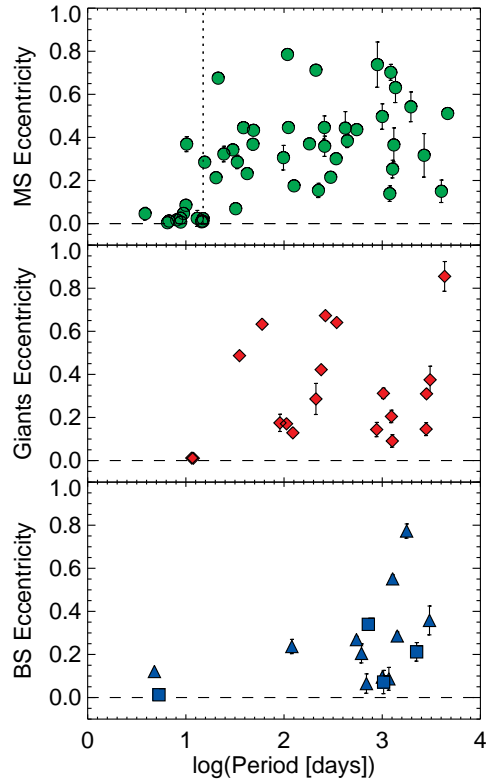}
\caption{\footnotesize Eccentricity plotted against the logarithm of the period for the main-sequence (top), giant 
(middle) and blue straggler (bottom) populations.  For the main-sequence binaries, we plot the tidal circularization
cutoff period $P_{circ}$ = 15.0 days \citep{mat04} with the dotted line.  For the blue straggler binaries, we identify the core (triangles)
and halo (squares) populations.}
\label{elogP}
\end{figure}

Tidal circularization for the NGC 188 MS binaries has been well studied by \citet{mat04} and \citet{mei05}.  
Here we reintroduce the eccentricity versus log period ($e - \log(P)$) diagram to present our new data, and to compare the 
MS, giant and BS distributions.  We show the diagram in Figure~\ref{elogP} for the MS (top), giant (middle) and 
BS (bottom) populations.  Here, as for all such figures in this paper, we differentiate the MS, giant and BS binaries
by plotting them in light gray, dark gray and black (green, red and blue in the online version), respectively.
For reference, in the MS plot we show 
with the dotted line the circularization cutoff period, $P_{circ}$ = 15.0 days, found by \citet{mat04} 
and consistent with the circularization period of 14.5 days found by \citet{mei05}.

The MS and giant $e - \log(P)$ diagrams are of a similar form to those of solar-type binaries in the Galactic 
field (R10) and open clusters of a wide range in age \citep[see][]{mei05}. The binaries are distributed across 
a large range in periods, with the short-period binaries ($P<P_{circ}$) having mainly circular orbits and 
longer-period binaries having a range in non-zero eccentricities.
A two-dimensional Kolmogorov-Smirnov (K-S) test (using the technique of \citealt{fas87}) comparing the MS and giant 
distributions reveals no statistically significant difference, although the expected lack of short-period orbits among the giants is present.  

However the BS binaries have a significantly different $e - \log(P)$ distribution than the MS
(as also discussed in \citealt{mat09}).
The majority of the BS binaries have periods on the order of 10$^3$ days.
A two-dimensions K-S test results in 99\% confidence that the 
MS and BS binaries are drawn from different parent populations.  
This concentration of BS binaries near periods of 10$^3$ days is also observed in the M67 BSs \citep{lat07} as well 
as in metal-poor field BS binaries \citep{car05}, and may provide significant insights into the origins of the BSs.

The mean eccentricity for long-period ($P > 500$ day) BS binaries is 0.28 $\pm$ 0.06, while
the mean eccentricity for similar MS binaries is significantly higher, at 0.44 $\pm$ 0.06 \citep{mat09}.  
Furthermore, three of these twelve long-period BS binaries have eccentricities within two standard deviations of zero, 
while none of the long-period MS binaries have such low eccentricities.

Additionally, in the bottom panel of Figure~\ref{elogP} we divide our BS binaries to compare the core and halo
BS populations (again divided at 10 arcmin from the cluster center).  We plot the eleven core BS binaries 
with triangles and the four halo BS binaries with squares.  We find no statistically 
significant distinction in either one- or two-dimensional K-S tests comparing the period and eccentricity distributions of 
these two populations, though these results are drawn from small sample sizes.

Finally, we note that none of our binaries with orbital solutions have periods of less than one day.  Indeed our shortest-period 
MS binary has a period of 3.8 days, and our shortest-period giant binary has a period of 11.5 days.
We can understand this lack of short-period giant binaries as an evolutionary effect due to increasing radius.
For two solar-mass stars, a period of 10 days corresponds to an orbital semi-major axis of $\sim$25 \Rsolar,
which is smaller than the radii of the NGC 188 giants \citep{van99,gir02}.

On the other hand the R10 period distribution predicts that we should observe $\sim$6-8 MS binaries with periods below 4 days.  
It is well known that NGC 188 contains a large number of 
W UMa systems \citep[e.g.][]{bal85,zha02,kaf03}.  There are four known W UMa's within 
our sample that have non-zero proper-motion memberships, each having a photometric period of less than one day.
We have been unable to obtain kinematic orbital solutions for these systems, as their rapid rotation hinders our 
ability to derive precise RVs.  This lack of detached short-period binaries combined with 
the high frequency of W UMa systems in NGC 188 suggests that we may be observing an evolutionary depletion of close binaries.
Perhaps this is a result of orbital angular momentum loss from magnetic braking in stellar winds that has slowly 
decreased the semi-major axes of the closest binaries \citep{bal85}.  
Alternatively, short-period binaries in triple systems may have been driven together through the Kozai 
mechanism and tidal friction \citep{egg06} or dynamical encounters.  
(We have not yet detected tertiary companions to any of the W UMa's in our sample, though kinematic
detection of tertiaries is impeded by the rapid rotation of the W UMa.)

\section{Eccentricity Distribution} \label{Sefreq}

\begin{figure}[!t]
%\plotone{f4.eps}
\plotone{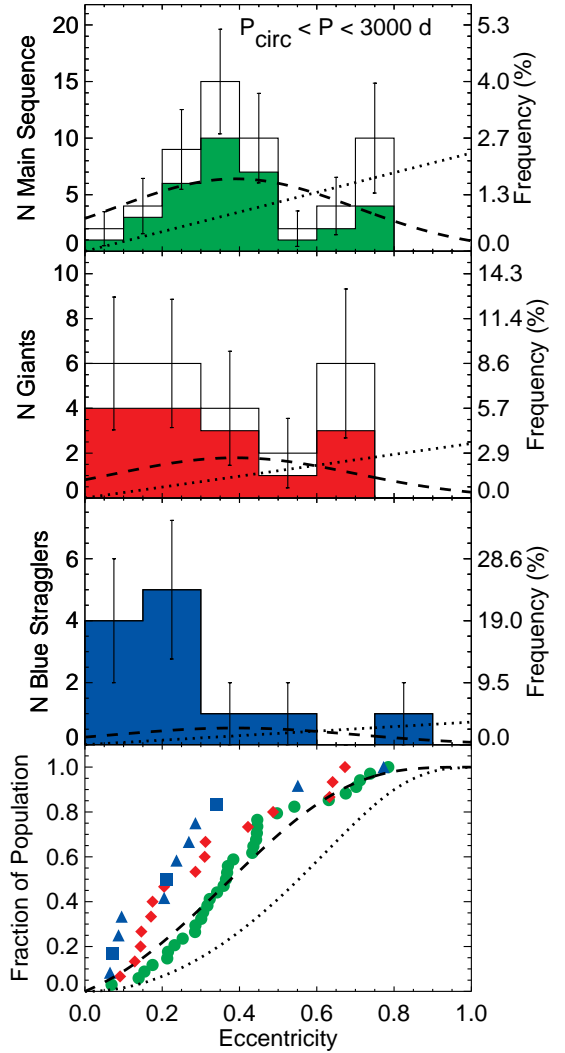}
\caption{\footnotesize 
Eccentricity distributions for NGC 188 binaries with $P_{circ} < P < 3000$ days.  
We show the observed main-sequence, giant and blue straggler binaries in light gray, dark gray and 
black (green, red and blue in the online version), respectively.
The top three plots show the distributions of eccentricity in histogram form.
We correct the main-sequence and giant distributions for our incompleteness and 
plot the corrected histograms with the solid lines above the observed distributions.
In each histogram plot, we also show a Gaussian fit to the \citet{rag10} field distribution
in the dashed line and a thermal distribution in the dotted line, both normalized to the respective sample size
using the field binary frequency within this period range of 11.6\%.
The bottom plot shows the cumulative distribution functions for the three populations in the same gray-scale (color in the online version) 
coding as for the histograms.  For clarity, we plot the main-sequence binaries with circles, giant binaries with diamonds, 
and we distinguish the core and halo blue straggler populations with the triangles and squares, respectively.
As a comparison, we also plot the field and thermal distributions corrected for our observational completeness in orbital solutions 
in the dashed and dotted lines, respectively.}
\label{efreq}
\end{figure}

Here we discuss the eccentricity distribution for binaries with periods 
between $P_{circ}$ and 3000 days, where $P_{circ} = 15.0$ days \citep{mat04}.  
In Figure~\ref{efreq} we plot the eccentricity distributions of NGC 188 binaries
in both histogram and cumulative distribution form.  We plot the detected binaries with orbital solutions 
in filled histograms.  The incompleteness-corrected distributions (for MS and giant binaries) 
are plotted with the solid-lined histograms above the observed distributions.
We choose to bin the MS data with bins of 0.1 and the giant and BS data 
with bins of 0.15 to reduce the scatter and larger errors associated with the relatively smaller number of 
giant and BS binaries. 
In the bottom panel of Figure~\ref{efreq}, we plot the cumulative distributions of these NGC 188 binary populations.

We correct the MS and giant distributions for incompleteness by applying the results of Section~\ref{incomp}.
(Again, we do not account for evolutionary effects in our Monte Carlo incompleteness study; e.g., 
to account for the increased radius of a giant star). 
As discussed in Section~\ref{incomp}, we do not attempt to correct the BS distributions for incompleteness 
as the masses and distributions of orbital parameters for these primaries are likely significantly different from the 
1~\Msolar~primaries and field dwarf binary orbital parameters used in our incompleteness analysis.
Again note that there are only six additional BSs in NGC 188, only one of which is detected as a binary. 
Thus any incompleteness correction for the BS binaries must be minimal.

We also show a Gaussian fit to the R10 field eccentricity distribution 
and a thermal distribution ($f(e) = 2e$) as the dashed and dotted lines, respectively, 
both normalized by the respective sample sizes using the R10 field binary frequency of 11.6\% in this period range.
In the cumulative distribution plot, we show the R10 and thermal distributions modified by our observational completeness
(in the corresponding line styles). 
To produce these curves we multiply the R10 and thermal distributions, respectively, by our incompleteness 
distributions in orbital solutions (from Figure~\ref{incfig}), effectively limiting these distributions to reflect our 
completeness level as a function of eccentricity.  These completeness-corrected distributions are then 
converted into the cumulative distributions shown in Figure~\ref{efreq}.

\begin{figure}[!t]
%\plotone{f5.eps}
\plotone{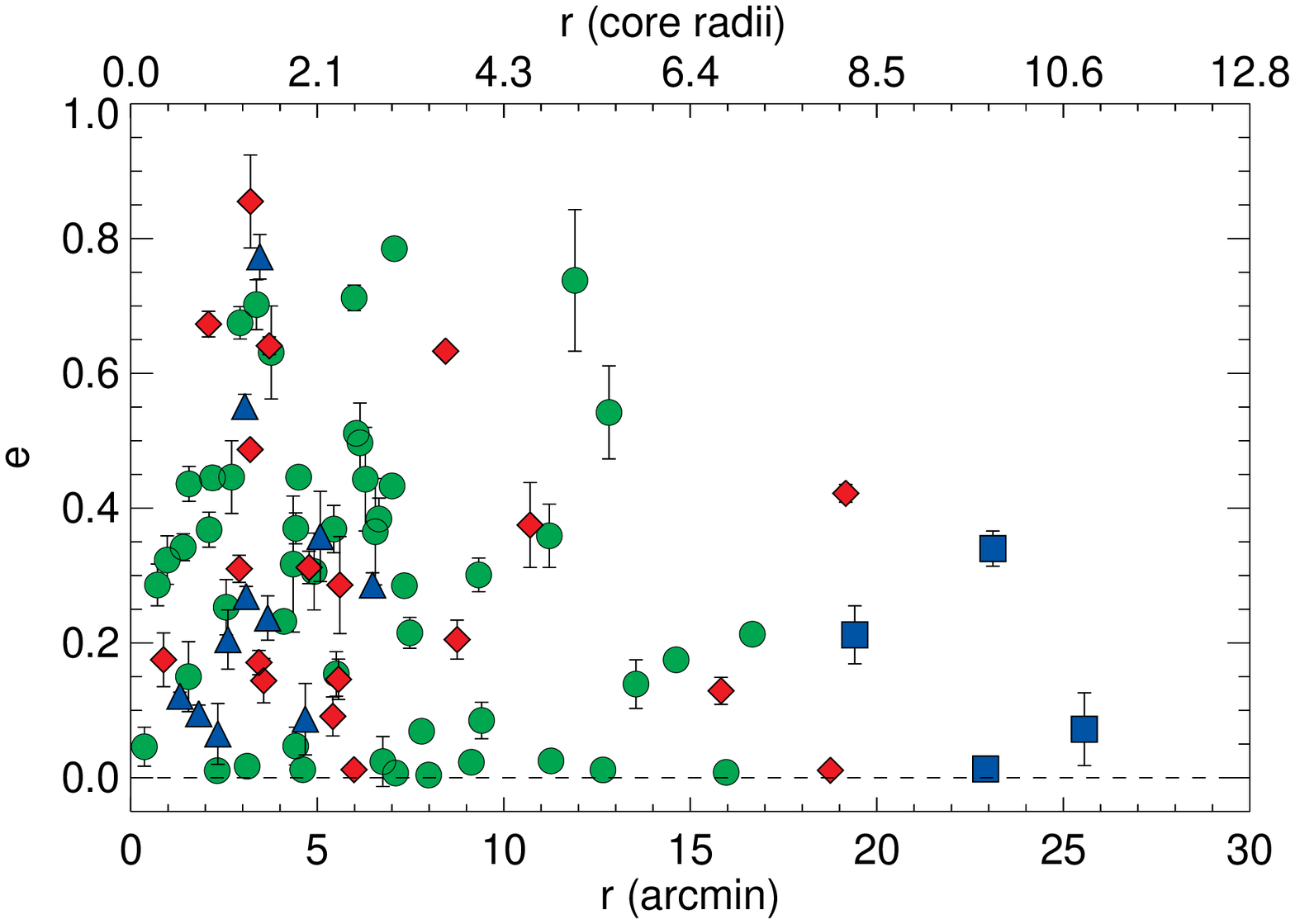}
\caption{\footnotesize Eccentricity as a function of projected radius from the cluster center.  
We plot the main-sequence binaries in light-gray circles, giant binaries in 
dark-gray diamonds and blue straggler binaries in black triangles (core) and squares (halo).
(We use green, red and blue for the main-sequence, giant and blue straggler binaries, respectively, 
in the online version.)
Note that we have not derived orbits for any binaries with $e > 0.5$ outside of 5 core radii from the cluster center.
When we divide the NGC 188 binaries at 5 core radii ($\sim13$ arcminutes), we find the inner binaries 
to have a distribution shifted to higher eccentricities at the 97\% confidence level.}
\label{evr}
\end{figure}

The NGC 188 MS eccentricity distribution has a roughly Gaussian form centered on $e = 0.35$.
We find no distinction between the MS NGC 188 eccentricity distribution and either the Gaussian fit to the R10 eccentricity distribution 
or a flat eccentricity distribution.
The NGC 188 MS eccentricity distribution also agrees 
with that of similar samples in open clusters of a wide range in age.  The mean eccentricity of similar binaries in 
both M67 (4 Gyr) and the Pleiades (150 Myr) is $\sim$0.4 \citep{mat90,mer92}.  Likewise, the eccentricity distribution 
of M35 ($\sim$150 Myr) is well approximated by a Gaussian with a mean eccentricity of $\sim$ 0.3 - 0.4 \citep{mat08}. 
Conversely a K-S test shows that the NGC 188 MS eccentricity distribution is not drawn from a thermal distribution at the
$>$99\% confidence level.

We find no statistically significant distinction between the giant and MS eccentricity distributions.  However a K-S test
comparing the BS and MS eccentricity shows the two distributions to be distinct at the 97\% confidence level, with the 
BS distribution shifted to lower eccentricities.  This result agrees with that found by \citet{mat09} and also 
discussed in Section~\ref{SelogP}, where we show that the BS binaries with $P>500$ days have significantly lower eccentricities
than MS binaries of similar periods.

We also investigate the dependence of eccentricity on projected radius from the cluster center.
In Figure~\ref{evr} we show all binaries with orbital solutions, no longer limiting by period.
As is clear from this figure,  we have not derived orbital solutions for any high eccentricity 
($e > 0.5$) binaries at $r \gtrsim 5$ core radii ($r \gtrsim 13$ arcminutes), while we do have such binaries in the 
inner-cluster region.
If we divide our sample of all binaries at 5 core radii, we find the 
inner binaries to have a mean eccentricity of 0.31~$\pm$~0.03, and the outer binaries to have a 
mean eccentricity of 0.16~$\pm$~0.04.
These two mean values can be distinguished at the 97\% confidence level by a Student's T-test.
Likewise a K-S test shows that the inner binaries have a distribution shifted to higher eccentricities at 
the 97\% confidence level.
This distinction may be the result of dynamical encounters that are expected to occur most frequency in the denser core of a cluster.
In star clusters, fly-by interactions are the most common type of encounter
\citep{bac96}, and can perturb a binary causing an increase in orbital eccentricity \citep{heg96}.  
Furthermore close three- and four-body interactions involving lower eccentricity binaries have a high likelihood of 
producing high-eccentricity products \citep{fre04,gie03}.  
However, we caution the reader that our completeness in orbital 
solutions drops off dramatically for very high eccentricity binaries (Figure~\ref{incfig}).

\begin{figure}[!t]
%\plotone{f6.eps}
\plotone{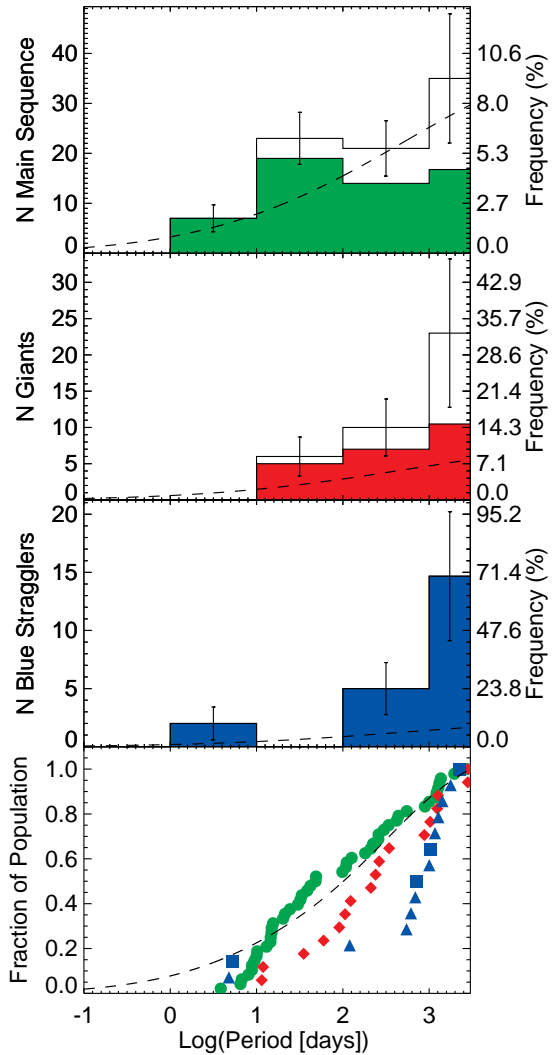}
\caption{\footnotesize 
Period distributions for the NGC 188 binaries with $P < 3000$ days.
We show the observed main-sequence, giant and blue straggler binaries in light gray, dark gray and black (green, red and blue in the online version),
respectively.
The top three plots show the period distributions in histogram form, while the bottom plot shows the 
cumulative distributions. The last bin extends to periods of 3000 days and is normalized to reflect the different bin size. 
We correct the main-sequence and giant binary period distributions for incompleteness and 
plot the corrected histograms with the solid line, above the observed distributions.  In each histogram plot, 
we also show the \citet{rag10} Galactic field log-normal distribution (dashed line) 
normalized to the respective sample sizes using the field binary frequency within this period range of 14.1\%.
In the bottom plot, we show the cumulative distributions of our three NGC 188 samples in the 
same gray-scale (color in the online version) coding as for the histograms.  Again, we plot the
main-sequence binaries in circles, giant binaries in diamonds and blue stragglers binaries in triangles (core) and squares (halo).
Here we plot the field distribution corrected for our 
observational completeness in orbital solutions with the dashed line.}
\label{Pfreq}
\end{figure}

\pagebreak

\section{Period Distribution} \label{SPfreq}

\begin{figure}[!t]
%\plotone{f7.eps}
\plotone{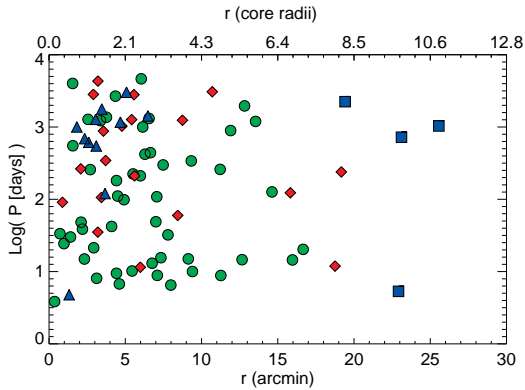}
\caption{\footnotesize The logarithm of the period as a function of projected radius from the cluster center.
We plot the main-sequence binaries in light-gray circles, giant binaries in 
dark-gray diamonds and blue straggler binaries in black triangles (core) and squares (halo).
(We use green, red and blue for the main-sequence, giant and blue straggler binaries, respectively, 
in the online version.)
We find no trend in period with radius.  (Note the errors in period, when 
plotted in the log, lie within the symbols used for plotting the data points.)}
\label{Pvr}
\end{figure}

In Figure~\ref{Pfreq} we plot our observed MS, giant and BS period distributions 
in both histogram and cumulative distribution form for binaries with $P < 3000$ days.  
Additionally we plot the R10 period distribution (dashed line) normalized to the respective sample sizes 
using the R10 binary frequency within this period range of 14.1\%.
For the MS and giant distributions we use our incompleteness study of Section~\ref{incomp} to correct our 
observed values.
In the bottom panel of Figure~\ref{Pfreq}, we plot the cumulative distributions of the NGC 188 binaries as compared 
to the R10 distribution modified by our observational completeness in the same manner as in Figure~\ref{efreq}.

The incompleteness-corrected NGC 188 MS log-period distribution rises towards our completeness limit,
in good agreement with the log-normal distribution found by R10 for solar-type field binaries. 
A K-S test comparing the MS NGC 188 and R10 cumulative distributions results in a 67\% probability that they are drawn 
from different parent distributions. This value is as large as 67\% due to the lack of observed binaries with 
periods less than $\sim$4 days, as discussed in Section~\ref{SelogP}.
Thus we find that the MS binary period distribution is 
indistinguishable from the R10 field distribution over our period range.  

A similar analysis 
returns $>$99\% confidence that the present day NGC 188 MS binary population is \textit{not} drawn from a flat 
distribution in log($P$), a common choice for initializing binaries in $N$-body models.
Additionally the flat period distribution is not supported by observations of the young 
($\sim$150 Myr) open cluster M35 \citep{mat08}.

A K-S test comparing the cumulative distributions of the MS and BS binary populations results in $>$99\% 
confidence that these two samples are drawn from distinct parent populations \citep{mat09}.  
The majority of the BS binaries are found at periods near 1000 days, as is also clear from Figure~\ref{elogP}.

A similar test comparing the giant and MS binaries results in an 88\% probability that they are drawn from 
different parent populations.  This marginal distinction
between the giants and MS binaries is partly due to the lack of short-period giant binaries.
This loss of short-period binaries is expected as the primary evolves off the MS
and expands to become a red giant.
If we exclude MS binaries with $P < 11.5$ days (the minimum orbital period found amongst the giant binaries in NGC 188),
the probability that the MS and giant distributions are different is reduced to $<$50\%.

We find no dependence in period with projected radius from the cluster center (Figure~\ref{Pvr}).
The core and halo BS binary populations are seen clearly in Figure~\ref{Pvr}.  There is no 
distinction between the period distributions of these two subsets of the BSs, even to the extent 
of each having one short-period SB2 BS binary.

\section{Solar-Type Main-Sequence Secondary-Mass and Mass-Ratio Distributions for Hard Binaries} \label{Sqm2freq} 

The secondary-mass and mass-ratio distributions for most binary populations are less well
known than, for instance, the period and eccentricity distributions.  Kinematic orbital solutions alone do not 
reveal the component masses for spectroscopic binaries.  For SB2 binaries we derive the mass ratio directly; 
however the individual masses obtained from the kinematic orbital solutions are dependent on the inclination angle.  
For single-lined (SB1) binaries, we only derive the mass function 
in which the primary and secondary masses are entangled and also dependent on the inclination.  

\begin{figure*}[!t]
\begin{center}
\epsscale{1.0}
%\plottwo{f8a.eps}{f8b.eps}
\plottwo{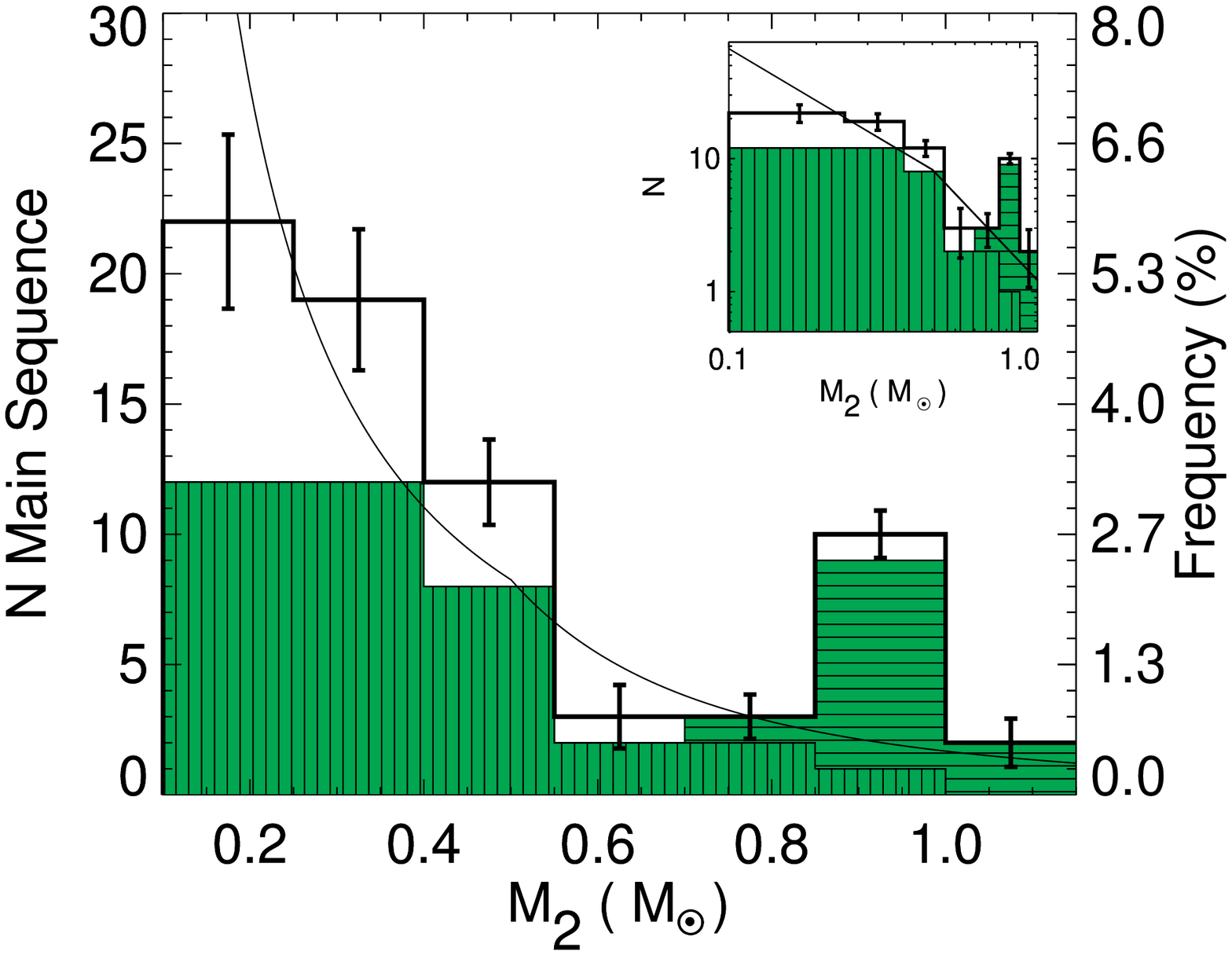}{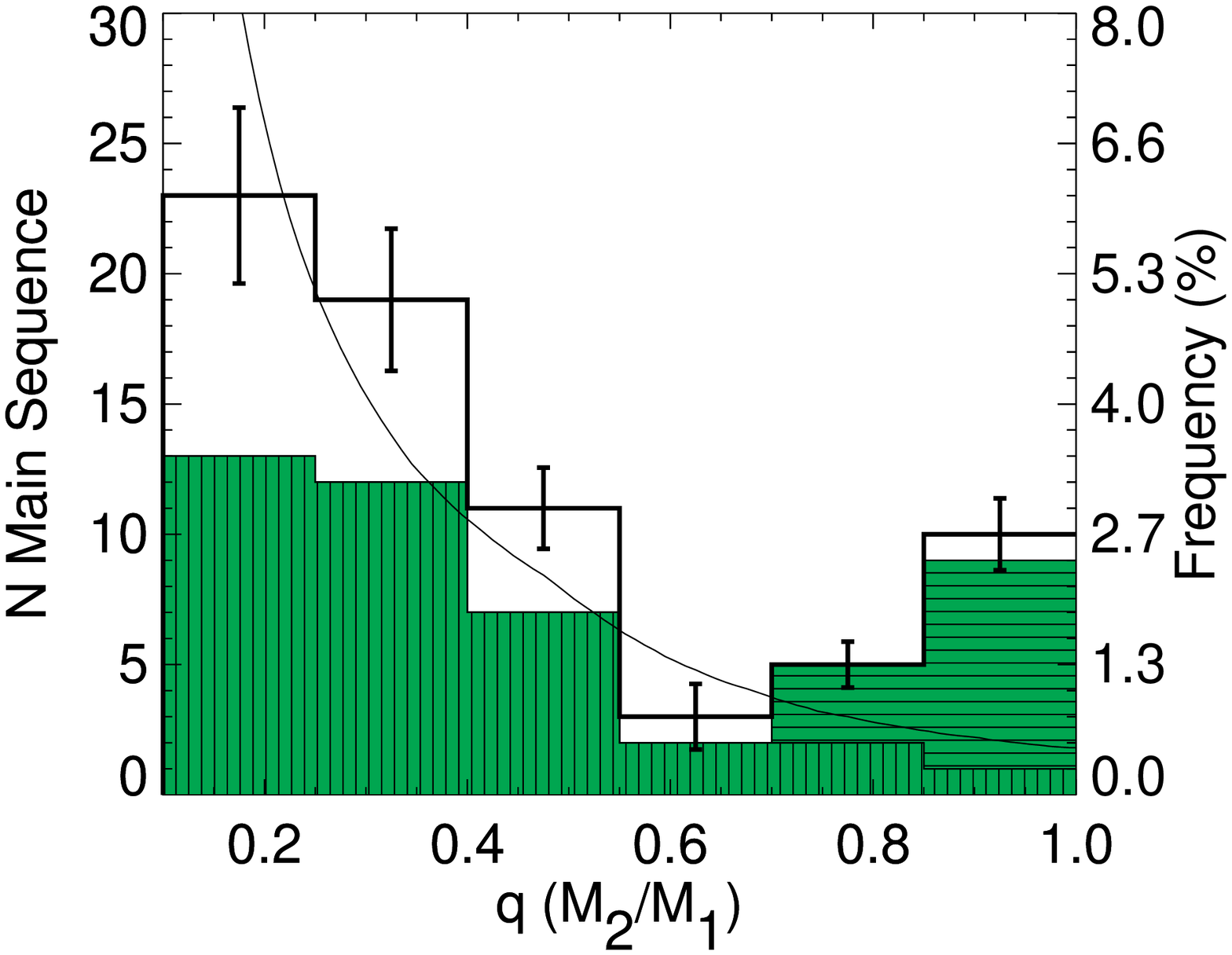}
\caption{\footnotesize Secondary-mass (left) and mass-ratio (right) distributions for the NGC 188 
solar-type main-sequence hard binaries.  The detected binaries with orbital solutions are plotted 
in the filled histograms, with SB2s in the horizontally-hatched histograms and SB1s in the vertically-hatched histograms.
For the SB1 binaries we utilize the statistical inversion technique of \citet{maz92} to derive
the secondary-mass and mass-ratio distributions from the observed mass functions.  
The incompleteness-corrected distributions are plotted as the thick solid lines above the observed distributions.
The error bars are derived through a Monte Carlo analysis utilizing the \cite{maz92} algorithm
to convert the Poisson errors on the mass function distribution into uncertainties on the secondary-mass and 
mass-ratio distributions, respectively, and we show the 95\% confidence intervals.
For comparison, in the secondary-mass panel we show the \citet{kro01a} IMF,  and 
in the mass-ratio panel we plot the distribution obtained from choosing 
random partners from this IMF, both with the solid curves.
The curves are normalized to contain the same number of binaries as in our incompleteness-corrected NGC 188 sample.
Finally we plot the secondary-mass distribution in the more familiar log-log format in the upper-right inset.
We show this figure for illustrative purposes, but focus our statistical analysis on the observed mass-function distribution
shown in Figure~\ref{fmfig}.
}
\label{qm2freq}
\end{center}
\epsscale{1.0}
\end{figure*}

In Paper 2 we used a photometric deconvolution technique to estimate primary  
masses for all of the MS binaries with orbital solutions.
Here we use these primary mass estimates and the dynamical mass functions within a statistical algorithm 
\citep{maz92} to derive the secondary-mass and mass-ratio distributions.
In Figure~\ref{qm2freq} we plot the resulting MS mass-ratio and secondary-mass distributions for the NGC 188 binaries. 
As in previous figures, all detected binaries with orbital solutions are shown in the filled histograms, 
while the incompleteness corrected distributions 
are plotted in the thick solid lines above the observed distributions.  
The horizontally-hatched regions in Figure~\ref{qm2freq} show the SB2 binaries and the vertically-hatched
regions show the SB1 binaries.

\begin{figure}[!t]
\begin{center}
\plotone{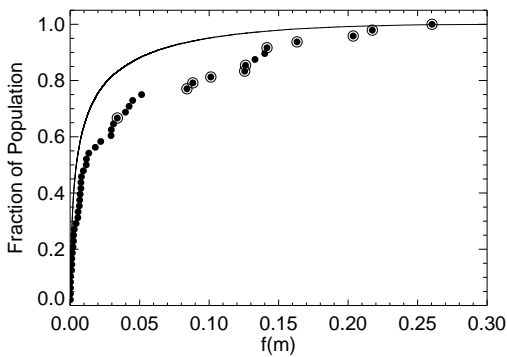}
\caption{\footnotesize Cumulative mass-function distribution for the NGC 188 solar-type 
main-sequence binaries.  We show the observed mass functions with the black points, and circle the 
double-lined binaries (for reference).  With the solid line we plot the mass function 
distribution derived by drawing companions from the \citet{kro01a} IMF.
To derive this curve we assume the inclination angles are distributed isotropically and draw primary masses 
from a Gaussian fit to the observed NGC 188 main-sequence primary-mass distribution.  This theoretical distribution is then corrected 
for our incompleteness in orbital solutions (out to $P=3000$ days) in the same manner as Figures~\ref{efreq} and \ref{Pfreq}.
A K-S test shows that the observed mass-function distribution is statistically indistinguishable from the IMF.
}
\label{fmfig}
\end{center}
\end{figure}

For the SB1 binaries we use the iterative algorithm of \citet{maz92} to convert our distributions of mass functions into distributions
of mass ratio and secondary mass, using our primary mass estimates and assuming the inclination angles to be randomly 
distributed.  Note that this technique (as well as other statistical inversion methods) does not reveal the secondary mass
or mass ratio for a given binary, but instead derives the respective distributions for a sample of binaries.

We correct the SB1 binaries for our incompleteness using the analysis of Section~\ref{incomp} after applying the 
statistical algorithm to derive the observed distributions.  We do not use 
this same incompleteness correction for the SB2 binaries as these systems (once detected) are given the highest priority 
in our observations.  As such, we have only one confirmed SB2 binary without an orbital solution (ID 9119), 
and we add in this binary by hand.  
The turnoff in NGC 188 is at $\sim$1.1 \Msolar, and our incompleteness correction covers binaries with $P < 3000$ days 
that have stellar companions down to 0.08 \Msolar~(excluding brown dwarfs and other sub-stellar objects).

In the secondary-mass plot, we also show the ``universal IMF'' from \citet{kro01a} with the solid curve.
In the mass-ratio plot, we show the distribution obtained from choosing 
random partners from the IMF in the solid curve, limiting the primary to be within our 
observable mass range and with the mass ratio always less than one.
The curves are normalized to contain the same number of binaries as in our incompleteness corrected NGC 188 sample.

The \citet{kro01a} IMF was not intentionally derived to model secondary stars in binaries.
Still the shape of the MS secondary-mass distribution is consistent with the IMF in that it rises towards 
lower masses.  Likewise the mass-ratio distribution shows a similar form to the curve derived by choosing random partners
from this IMF.  

We also observe an excess of binaries with masses between 0.85 and 1.0 \Msolar~(and mass ratios $>0.9$) as compared to the IMF.  
This excess of near equal-mass binaries remains if we instead use the mass functions for all binaries (including the SB2s) and derive 
the secondary-mass and mass-ratio distributions for the full binary population with the \citet{maz92} algorithm; therefore this result 
is not sensitive to the method used here.
This may be analogous to the abundance of ``twins'' observed 
in some studies, generally at shorter periods \citep[e.g.][R10]{tok00,fis05}. We have investigated 
the mass-ratio distribution as a function of period, but find that the decreased sample sizes make these
results highly uncertain.  

To check formally for consistency between the observed companion masses and the IMF
we turn to the observed mass functions. In Figure~\ref{fmfig}, we compare the observed cumulative mass-function
distribution (black points, with SB2s circled) to a theoretical distribution (solid line) derived by drawing companions from 
the IMF, primary masses from a Gaussian fit to our observed NGC 188 MS primary mass distribution, and assuming an isotropic inclination 
distribution.
We then correct this theoretical distribution for our incompleteness in orbital solutions using the same method as in
Figures~\ref{efreq} and \ref{Pfreq}.  A K-S test shows that we cannot formally distinguish the observed 
mass-function distribution from the IMF.  (We also note that the mass-function distribution cannot formally be distinguished from 
a distribution derived by choosing binaries from a uniform mass-ratio distribution.)

\begin{figure*}[!t]
\begin{center}
\epsscale{1.0}
%\plottwo{f10a.eps}{f10b.eps}
\plottwo{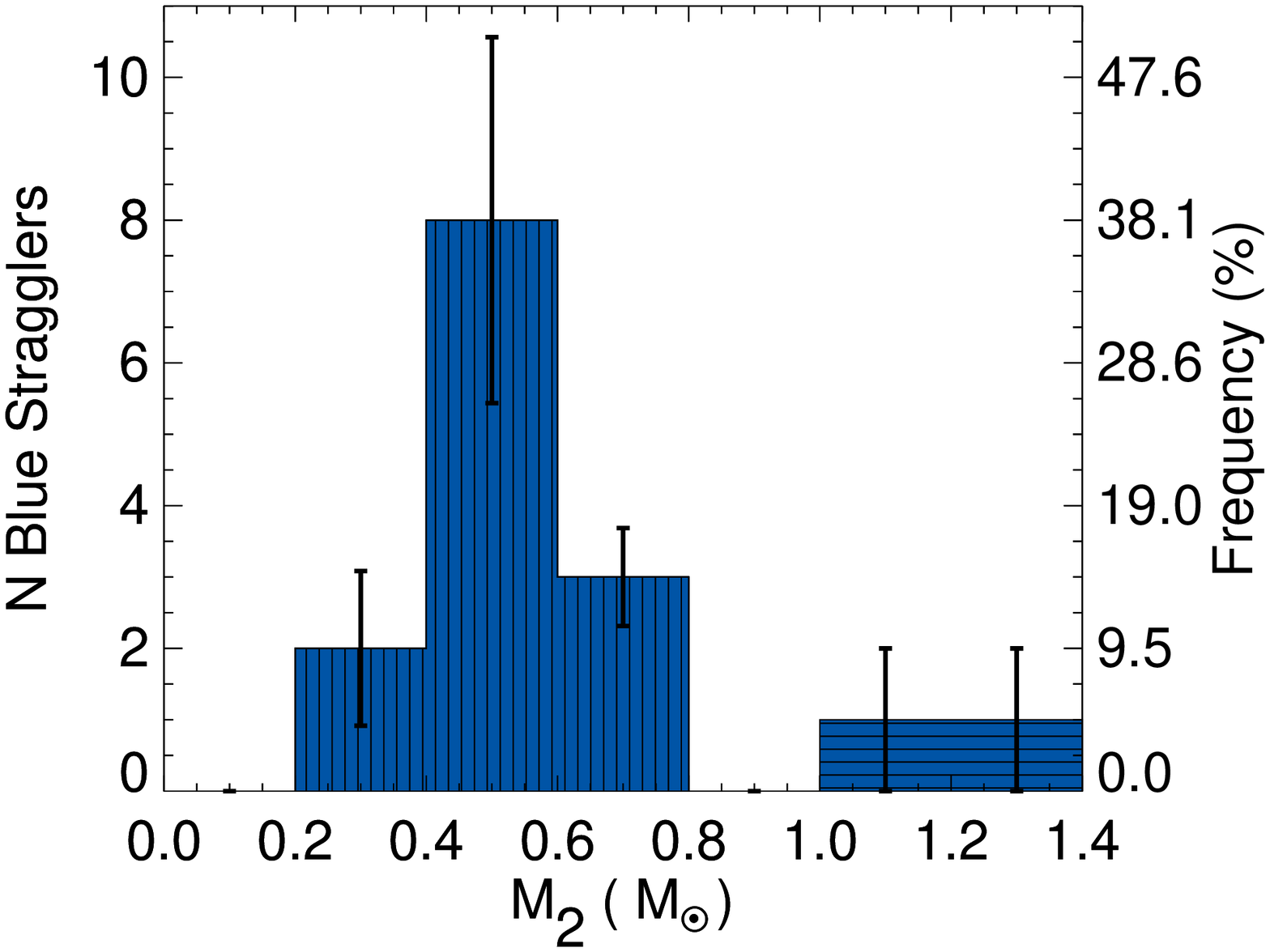}{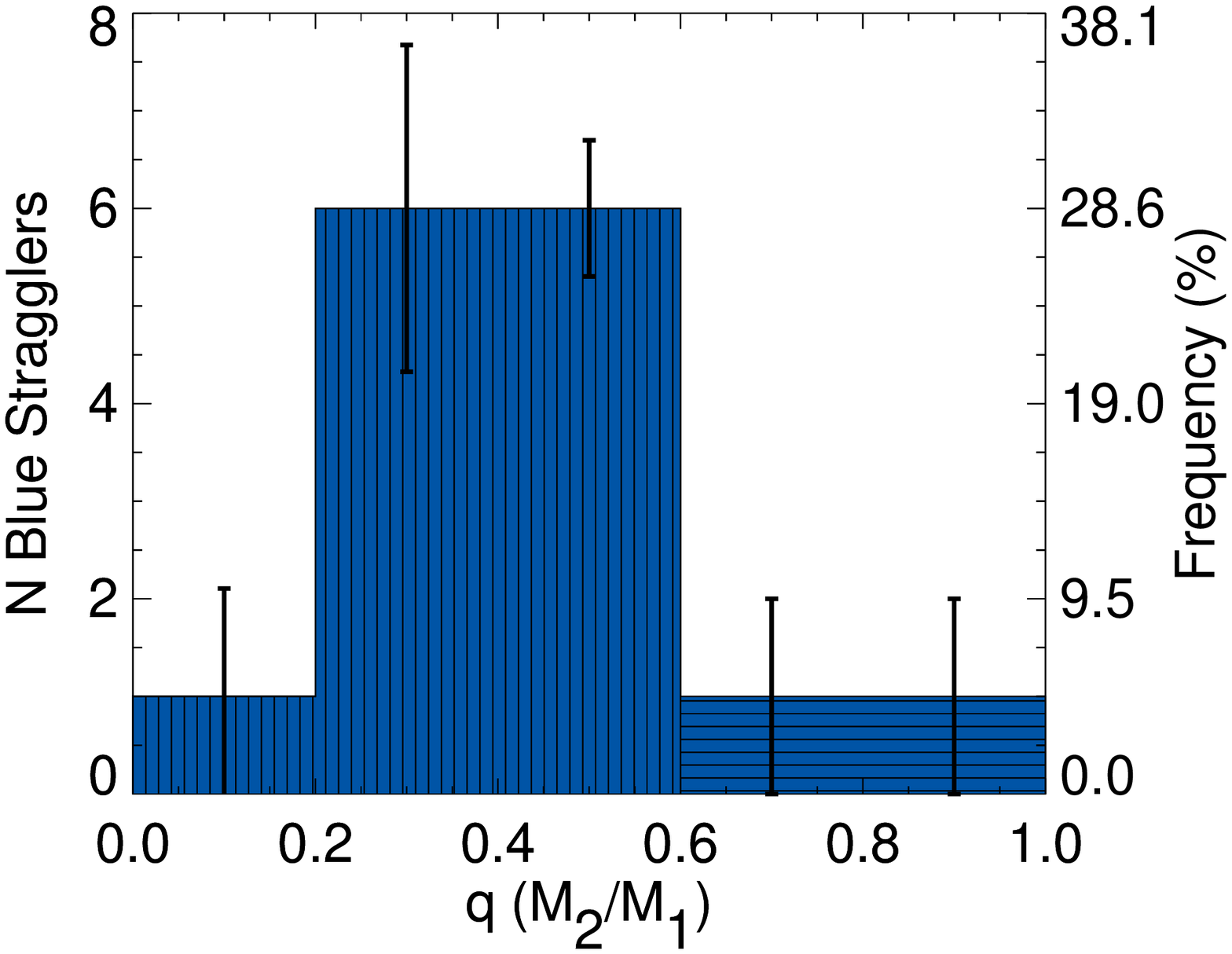}
\caption{\footnotesize Secondary-mass (left) and mass-ratio (right) distributions for the NGC 188 
blue straggler binaries.  The SB2s are plotted with the horizontally-hatched histograms and the SB1s are plotted 
with the vertically-hatched histograms.  For SB1s, the primary masses are estimated in relation to 
evolutionary tracks of \citet{mar08}, and we then use the \citet{maz92} algorithm to derive the secondary-mass 
and mass-ratio distributions from the observed mass functions.  The masses for the two SB2 blue stragglers are discussed in 
detail in Section~\ref{Sqm2freqBS}, and see also \citet{mat09}. The uncertainties shown here
are derived in the same manner as in Figure~\ref{qm2freq}. }
\label{qm2freqBS}
\end{center}
\epsscale{1.0}
\end{figure*}

There is some disagreement in the literature about the secondary-mass and mass-ratio distributions for field dwarfs.
R10 find their sample of field dwarfs to have a uniform mass-ratio distribution between $0.2 < q < 0.95$, with evidence
for an abundance of twins at $q > 0.95$.   \citet{maz03} find a mass-ratio distribution that is formally consistent with a uniform distribution.
They also note a possible rise in the distribution towards lower mass-ratio values, but with large uncertainties.
\citet{duq91} and \citet{gol03} find distributions that rise towards lower-mass secondaries.
(Strictly \cite{gol03} find this result for disk binaries, and not for the halo binaries.)
We conclude that the companions to the NGC 188 MS binaries are consistent with 
those of the field dwarf binaries, but given the uncertainty in the literature about the field population this conclusion lacks precision.

\section{Blue Straggler Secondary-Mass and Mass-Ratio Distributions} \label{Sqm2freqBS}

The masses and evolutionary states of the companion stars to the BSs in binaries 
are a powerful tool to investigate the origins of these systems.
Knowledge of whether the secondaries are normal MS stars or white dwarfs (WDs) can be particularly useful in 
distinguishing between possible formation mechanisms \citep{gel11}.  
For instance, the Case B or Case C mass transfer mechanisms
predict a BS binary with a WD companion \citep{mcc64,pac71}, while a merger of an inner binary of a triple system
\citep{per09} or a stellar collision that leaves a bound companion \citep{hur05} would more likely produce MS secondaries.

\begin{figure}[!t]
\begin{center}
\plotone{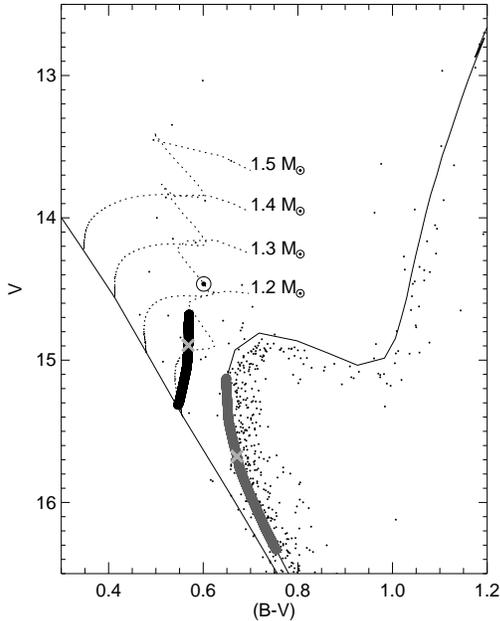}
\caption{\footnotesize Color-magnitude diagram showing the loci of possible locations for the components of the blue straggler 
binary ID 5078.  Cluster members are shown in the small black points, and 5078 is circled.  We also show a ZAMS and a 7 Gyr 
isochrone in the thin solid lines, and evolutionary tracks for 1.2~\Msolar, 1.3~\Msolar, 1.4~\Msolar~and 1.5~\Msolar~stars in the 
dotted lines, respectively \citep{mar08}. The thick gray line shows the locus of potential secondary stars, and the thick black 
line shows the locus of potential primary stars.  
The derived effective temperature of 5850 K for the secondary implies a mass
of 1.02~\Msolar; we plot the location of this secondary star, and the
location of the associated primary, with the light-gray crosses. However, given the kinematic mass ratio of $q=0.678\pm0.009$
and this secondary mass, the blue straggler mass would be 1.5~\Msolar.  Yet the blue straggler is significantly less luminous than
a normal 1.5~\Msolar~star at any point in its evolution.
}
\label{CMD5078}
\end{center}
\end{figure}

In order to derive the observed BS secondary-mass and mass-ratio distributions, we must first estimate the 
BS primary masses. To do so we compare the observed magnitudes and colors of the BSs to a coarse grid of 
evolutionary tracks \citep{mar08} extending from the ZAMS to 7 Gyr, and for stars with masses ranging from the 
cluster turnoff mass of 1.1~\Msolar~to twice this mass (in steps of 0.1~\Msolar).
For the SB1 BSs, the secondaries are at least 2.5 magnitudes (in $V$) fainter than the BS primaries (Paper 1). 
For a given BS, deconvolving a secondary star at this limiting magnitude from the combined light would result 
in a primary of $\lesssim 0.1$ $V$ magnitudes fainter than the combined $V$ magnitude.  This minor difference in 
$V$ magnitude is below the mass resolution of the evolutionary tracks (e.g., see Figure~\ref{CMD5078}).
Therefore we do not attempt to deconvolve the secondary light from these BSs, and estimate their masses based on 
the proximity of the combined magnitude and color to the evolutionary tracks.
We find the NGC 188 SB1 BSs to have masses between 1.2 - 1.6 \Msolar.  

We then use the \citet{maz92} algorithm in the same manner as for the MS binaries to derive the BS binary
secondary-mass and mass-ratio distributions shown in Figure~\ref{qm2freqBS} 
(with SB1s in the vertically cross-hatched histograms and SB2s in the horizontally cross-hatched histograms).
The BS SB1 secondary-mass distribution is narrow and peaks near a mass of $\sim$0.5~\Msolar. 
This result, when considered in light of the periods near 1000 days and low eccentricities, suggests an origin through 
Case C mass transfer, which predicts 
carbon-oxygen WD companions of masses between $\sim$0.5 - 0.6 \Msolar~with orbital periods of order 1000 days \citep{mcc64,pac71,hur02,che08}.
However the secondary-mass distribution alone does not rule out MS companions, as we would not detect the flux from 
a 0.5~\Msolar~MS (or WD) star in our BS spectra due to the large difference in luminosities. 
We discuss this result in detail in \cite{gel11} and build upon these results in Section~\ref{Hcomp}.

For the two short-period SB2 BS binaries, the secondary adds significant light to the system, and thus for each SB2 BS
we deconvolve the secondary from the combined light before estimating the primary mass. Additionally 
we utilize a grid of synthetic spectra to estimate the effective temperatures of both components of these
systems, which helps to further constrain the masses.  We discuss these results briefly in \citet{mat09} and expand upon them here.

BS ID 7782 has a kinematic mass ratio of $q=1.005\pm0.013$ (Paper 2). 
A spectral analysis yields a luminosity ratio\footnote{Due to an update in our TODCOR analysis we find a slightly different luminosity ratio 
from \citet{mat09}, who gave a value of $0.75 \pm 0.01$.} of $0.871 \pm  0.024$, 
and effective temperatures for the primary and secondary stars
of 6500 K and 6325 K, respectively.
Both stars are hotter than the MS turnoff effective temperature of 5900 K and more luminous than turnoff stars. 
Evidently 7782 comprises \textit{two} BSs.
The location of the system in the cluster halo suggests that a close dynamical encounter
was involved in the formation of the binary, perhaps the final of several encounters near the cluster core 
that exchanged the BSs into the system or formed one of them collisionally and ejected the binary into the cluster halo.

The differing temperatures and luminosities are unexpected given the nearly identical masses. 
Possibly the two BSs did not form at the same time and thus have different evolutionary ages. 
We are able to model the system with two 1.25 \Msolar~stars having an age difference of $\sim$2 Gyr \citep{mar08}, although this solution is not
unique given the present data. More detailed spectral analysis of the two stars is merited.

Nonetheless, given the large bin sizes used in constructing the secondary-mass
distribution (Figure~\ref{qm2freqBS}),
we can confidently place the secondary of 7782 in the 1.2 - 1.4 \Msolar~bin.

BS ID 5078 has a kinematic mass ratio of $q=0.678\pm0.009$ (Paper 2).  
From our analysis of the spectra, we derive an effective temperature for the primary of 6500 K and for the secondary of 
5850 K.  
The effective temperature of the secondary is consistent with a star on the upper MS.
Assuming the secondary is a MS star, we deconvolve the combined light using secondary light contributions derived 
from a 7 Gyr \citet{mar08} isochrone.  
We limit the possible secondaries such that the effective temperature is within 250 K ($\sim 2\sigma$) of 5850 K, and 
we limit the possible BS primaries such that the BS is on or redder than the ZAMS.
The resulting loci of possible CMD locations for the primary and secondary are shown in the thick black and gray lines 
in Figure~\ref{CMD5078}, respectively.

We find a range in possible masses for the secondary of 5078 of 0.94~\Msolar~to~1.08~\Msolar.
The nominal effective temperature for the secondary of 5850 K corresponds to a 1.02~\Msolar~MS star, and
we use this secondary mass in Figure~\ref{qm2freqBS}.
The location of the associated BS primary lies on or near a 1.2~\Msolar~evolutionary track.
We note though that the effective temperature of 6500 K
derived for the BS primary suggests a $(\bv)$ color that is $\sim$0.1 mag bluer than inferred from this analysis.

The range in secondary masses and the kinematic mass ratio yields a range for the BS primary star mass between
1.39~\Msolar~and~1.59~\Msolar. However, the locus of potential CMD locations for the primary clearly falls below such evolutionary
tracks, and even the combined light of 5078 is too faint. 
The standard stellar evolution tracks underestimate the dynamically informed mass of the BS primary star by 
15\% to 30\%.

This potential systematic error does not meaningfully effect the results shown in Figure~\ref{qm2freqBS}.
If we simply assign the same mass to each BS within the observed range of 
1.2 - 1.6 \Msolar, the resulting secondary-mass and mass-ratio distributions stay within the uncertainties 
on the distributions shown in Figure~\ref{qm2freqBS}.  Assigning all BSs a mass of 1.3 \Msolar, which is 
approximately the average BS mass that we derive, results in distributions that are nearly identical to those 
shown here.  If we take the extreme case of setting all BSs to twice the turnoff mass (2.2 \Msolar),  the general forms of the distributions 
remain the same, but the peak in the secondary-mass distribution shifts to the 0.6 - 0.8~\Msolar~bin.

Again we do not attempt to correct these BS distributions for our incompleteness and show only 
the BS binaries with orbital solutions. 
We remind the reader that there are only six additional NGC 188 BSs, only one of which is detected as a binary but does
not yet have an orbital solution.

\section{Comparison to a Sophisticated $N$-body Open Cluster Simulation} \label{Hcomp}

\subsection{The Simulation}

The open cluster $N$-body simulation of M67 by \citet{hur01,hur05} was able to track the evolution and dynamics of 
realistic initial numbers of single stars \textit{and} binaries, and included 
detailed prescriptions for mass loss, tidal processes, mass transfer, mergers and collisions.  
Through their model, Hurley et al.~were able to match the observed M67 color-magnitude diagram remarkably well.  
They produced similar populations of anomalous stars as are observed in M67, including BSs.
Importantly, they were able to track the dynamical evolution of 
these anomalous stars and pinpoint their origins with great detail, many of which involved interactions with 
binary stars. Simulations such as these have the power to revolutionize our understanding of the evolution of a binary 
population and the origins of anomalous stars, like BSs, in star clusters.  

Though Hurley et al.~designed their simulations to model M67, their results can also be compared to similar
open clusters, such as NGC 188.  Both NGC 188 and M67 are old open clusters at 4 and 7 Gyr respectively.
The current dynamical relaxation time for NGC 188 is 180 Myr \citep{chu10} and for M67 is 100 Myr \citep{mat86}.
As shown in Section~\ref{freq}, NGC 188 and M67 have comparable binary frequencies, and their binary orbital 
period and eccentricity distributions are of a similar form \citep{lat07}. 
Furthermore, the BS populations of these two clusters show strong similarities.  
\citet{lat96} observed $\sim$60\% of the M67 BS stars to be members of hard-binary systems, 
consistent with the 76~$\pm$~19~\% binary frequency amongst the NGC 188 BSs observed here.  Also, the M67 BS binaries show a 
similar tendency towards periods of order 1000 days \citep{lat96}.  
Thus, in this section we compare the NGC 188 hard-binary population to that of the \citet{hur05} simulation.

The simulation started with a 50\% binary frequency over all periods and masses, %(40\% for $P<10^4$ days), 
in which binary orbital elements were chosen from a flat period distribution and a thermal eccentricity distribution.  
The primary stars were chosen to follow a \citet{kro91} IMF, with masses $> 0.1$ \Msolar, and 
the mass ratios for binaries were chosen from a uniform distribution in $q$.  The binaries were
then modified according to the \citet{kro95b} tidal evolution prescription, and the resulting initial stellar population 
was evolved for 4 Gyr.  

We compare the MS and giant populations of the simulation at this final age to NGC 188.
We only analyze MS and giant stars that would be within our observable period, spatial and magnitude domain.
Specifically we consider only those binaries with $P < 10^4$ days to be detectable, and assume that we would only 
obtain orbital solutions for binaries with $P < 3000$ days. 
Our analysis only includes the inner 17 pc (radius in projection along the $z$ axis) of the cluster.
We use the NGC 188 distance modulus of $(M-m)_V$=11.44 and the reddening $E(\bv)$=0.09 \citep{sar99} 
to define the lower-mass (faint) limit of the MS.  The giants are all within our observable magnitude domain; thus we only 
limit the giants by radius from the cluster center.  

We choose not to impose magnitude or spatial limits on the BSs created within the simulation.
There are only four BSs created within the 
simulation throughout the entire cluster lifetime that at some moment reside outside of our observable spatial region.  
Each of these were formed near the core, and later ejected from the cluster, spending only one time step ($\sim$60 Myr) outside of 17 pc 
(in projection) from the cluster center before traveling beyond the cluster's tidal radius.

In the following analysis, we use an "integrated sample" of all BSs that existed between 2 and 4 Gyr.  The 
``integrated sample'' comprises the union of every BS at each time step in the simulation during this range in time. 
There are 57 individual BSs present at some point during this interval, while the integrated sample contains multiple 
realizations of these BSs over these 33 time steps.
In so doing, the integrated sample effectively weights results by the lifetime of each individual BS in a certain binary configuration.  
For example, there were 26 BSs created in binaries between 2 - 4 Gyr, 11 of which underwent an encounter after formation that 
significantly changed the orbital parameters.  Such changes in orbital parameters are accounted for, in the integrated 
sample, as separate systems with lifetimes that are equal to the duration at each orbital configuration.  

To close, Hurley et al.~did not have access to RV data as comprehensive as ours for either the M67 or NGC 188 binary 
population when constructing their simulation or analyzing their results.  Therefore, we are able to investigate the accuracy 
of their model both in more detail and, perhaps more importantly, in observational dimensions not available to them.
Many of the comparisons discussed next truly test \textit{a priori} predictions of their theory.

\begin{figure}[!t]
%\plotone{f12.eps}
\plotone{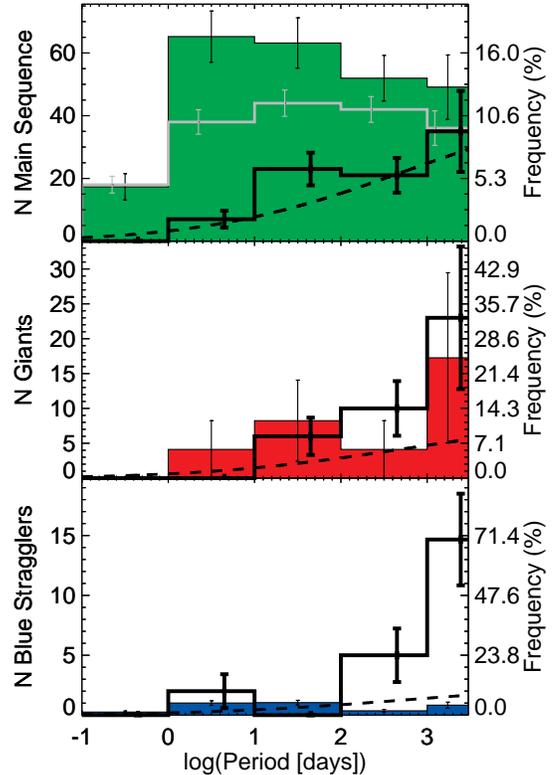}
\caption{\footnotesize Period distribution for binaries in the \citet{hur05} simulation.  
We compare the main-sequence (top), giant (middle) and blue straggler (bottom) populations from the 
simulation, using the 4 Gyr main-sequence and giant binaries and the integrated blue straggler sample.
Here we include all binaries with $P < 3000$ days, and the last bin is normalized to account for the different bin size.
In the main-sequence plot, we also show the initial period distribution of the simulation within the same mass, 
period and spatial domain with the gray line.   
All simulated distributions are normalized by the respective NGC 188 sample sizes, and we show the true frequencies on the right y-axes.
The respective NGC 188 distributions are plotted in the thick-lined histograms.
For comparison we also show the Galactic field period distribution of \citet{rag10} with the dashed lines,
normalized to the NGC 188 sample size using the field binary frequency within this period range.
The simulation contains a large overabundance of short-period main-sequence binaries as compared to observations of both NGC 188 and the field.
\label{HPfreq}
}
\end{figure}

\subsection{Comparison of the Main-Sequence Hard-Binary Populations in NGC 188 and in the Simulation} \label{ShMS}

\subsubsection{Frequency}

We find a MS solar-type hard-binary frequency at 4 Gyr in the simulation of 64~$\pm$~4~\% (230/361) for $P<10^4$ days.
(Here the uncertainty is the expected counting error given multiple realizations of the simulation.)
We specify the binaries defining this frequency in the same manner as for the MS NGC 188 binaries in Section~\ref{freq}.
The MS hard-binary frequency in the simulation has grown from its initial value of 47.9~$\pm$~2.5~\% 
(for the same mass, spatial and period domain).
This increase in binary frequency is a result of the preferential evaporation of the solar-type single stars 
over their higher total-mass solar-type binary counterparts.

The 4 Gyr MS solar-type binary frequency is twice that of the NGC 188 MS hard-binary frequency 
of 29~$\pm$~3~\%, and these two frequencies can be distinguished at the $>$99\% confidence level.
The simulated giant hard-binary frequency of 52~$\pm$~16~\% (11/21) is also higher than for the 
incompleteness-corrected NGC 188 giant hard-binary frequency of 
31~$\pm$~7~\%, although given the large uncertainties these two values are not distinct at high confidence.
These differences in binary frequency are directly linked to the high initial hard-binary frequency in the simulation,
and the small subsequent change in frequency during the cluster lifetime.

\begin{figure}[!t]
%\plotone{f13.eps}
\plotone{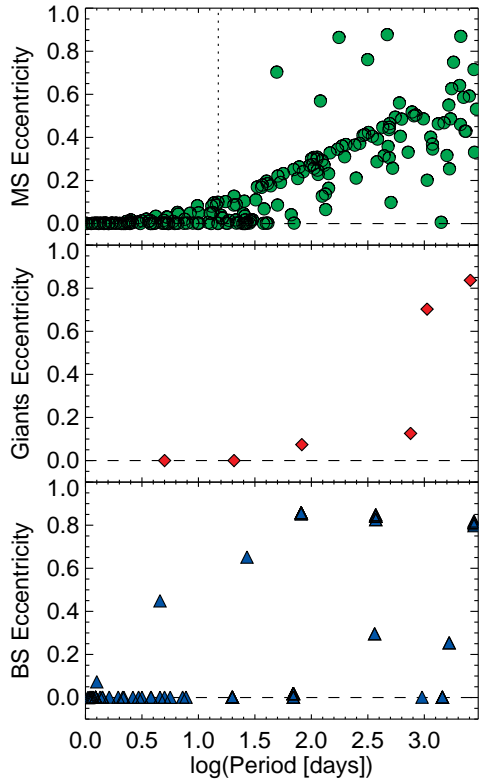}
\caption{\footnotesize $e - \log(P)$ diagram for binaries in the \citet{hur05} simulation.  We
compare the main-sequence (top), giant (middle) and blue straggler (bottom) populations from the simulation.
We show the 4 Gyr main-sequence and giant binaries.
For the blue stragglers, we show the integrated (2 - 4 Gyr) sample, where each point represents a given blue straggler binary configuration during 
one time step.  If a dynamical encounter alters the orbital period or eccentricity of a given blue straggler in a binary, the result is 
shown by individual (often partially overlapping) points at each orbital configuration, respectively. 
The circularization period $P_{circ} =15.0$ days \citep{mat04} is shown in the main-sequence $e - \log(P)$ plot by the dotted line.
Note the unphysical ``envelope'' to the main-sequence plot that is not seen in the NGC 188 observations (or those of other open clusters
of a wide range in age, \citealt{mei05}).
\label{HelogP}
}
\end{figure}

\begin{figure}[!t]
%\plotone{f14.eps}
\plotone{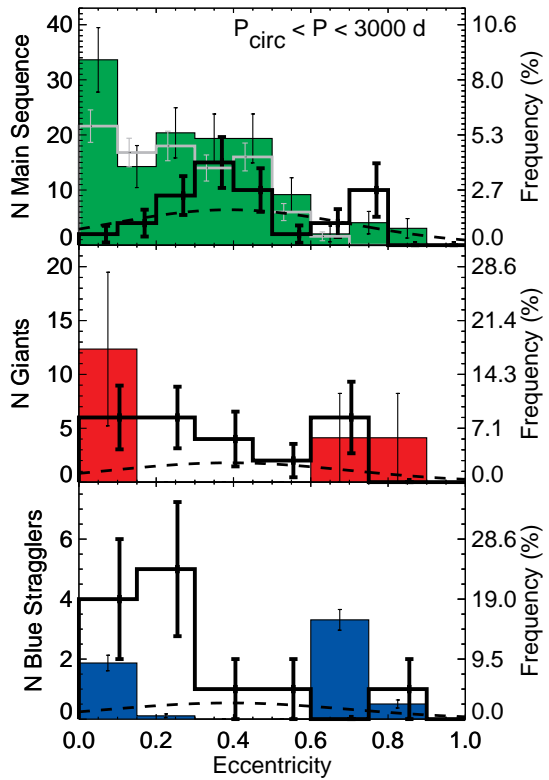}
\caption{\footnotesize Eccentricity distribution for binaries in the \citet{hur05} simulation.  We
compare the main-sequence (top), giant (middle) and blue straggler (bottom) populations from the 
simulation, using the 4 Gyr main-sequence and giant binaries and the integrated blue straggler sample.
Here we show only binaries with $P_{circ} < P < 3000$ days, where the circularization period $P_{circ} = 15.0$ days \citep{mat04}.
In the main-sequence plot, we also show the initial eccentricity distribution of the simulation within the same mass, 
period and spatial domain with the gray line.   
All simulated distributions are normalized by the respective NGC 188 sample sizes, and we show the true frequencies on the right y-axes.
The respective NGC 188 distributions are plotted in the thick-lined histograms.
Finally, we show the \citet{rag10} distribution in the dashed lines, normalized to the NGC 188 sample size using the 
field binary frequency within this period range.
\label{Hefreq}
}
\end{figure}

\subsubsection{Orbital Period}

In Figure~\ref{HPfreq} we plot the period distribution
for the simulation.  
We use the same gray-scale (color in the online version) coding for the simulated MS, giant and BS populations as we 
have done throughout the paper for NGC 188. %, but now corresponding to the populations from the simulation. 
We normalize the simulated output by the respective NGC 188 sample sizes
and show the true frequencies on the right y-axes.
For reference, 
we also include the NGC 188 distributions as the thick histograms and the normalized R10 field distributions as the dashed lines.
As in Figure~\ref{Pfreq} we show binaries out to $P = 3000$~days in Figure~\ref{HPfreq}, and normalize the last bin to 
account for the different bin size.

\begin{figure*}[!ht]
\begin{center}
%\plottwo{f15a.eps}{f15b.eps}
\plottwo{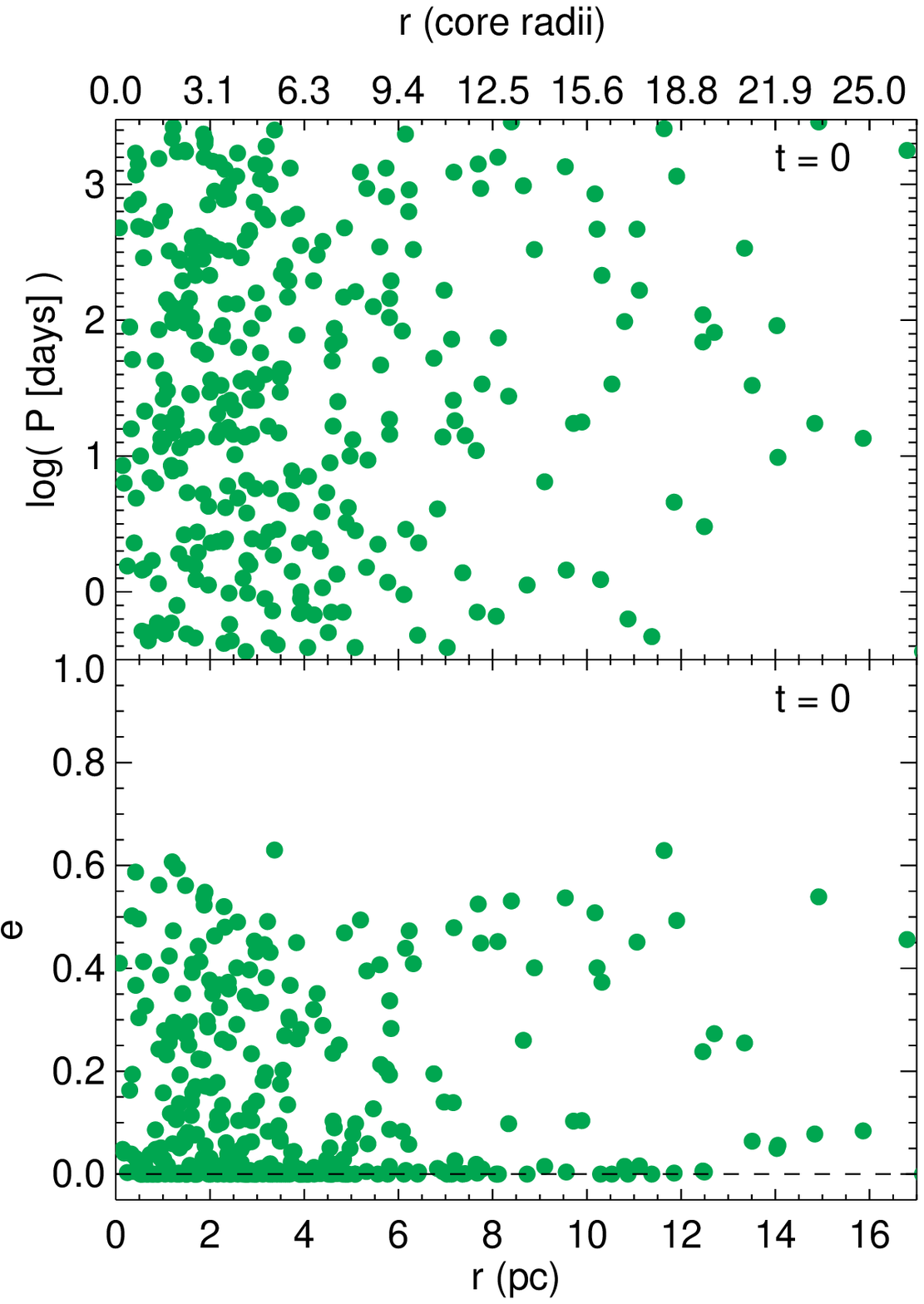}{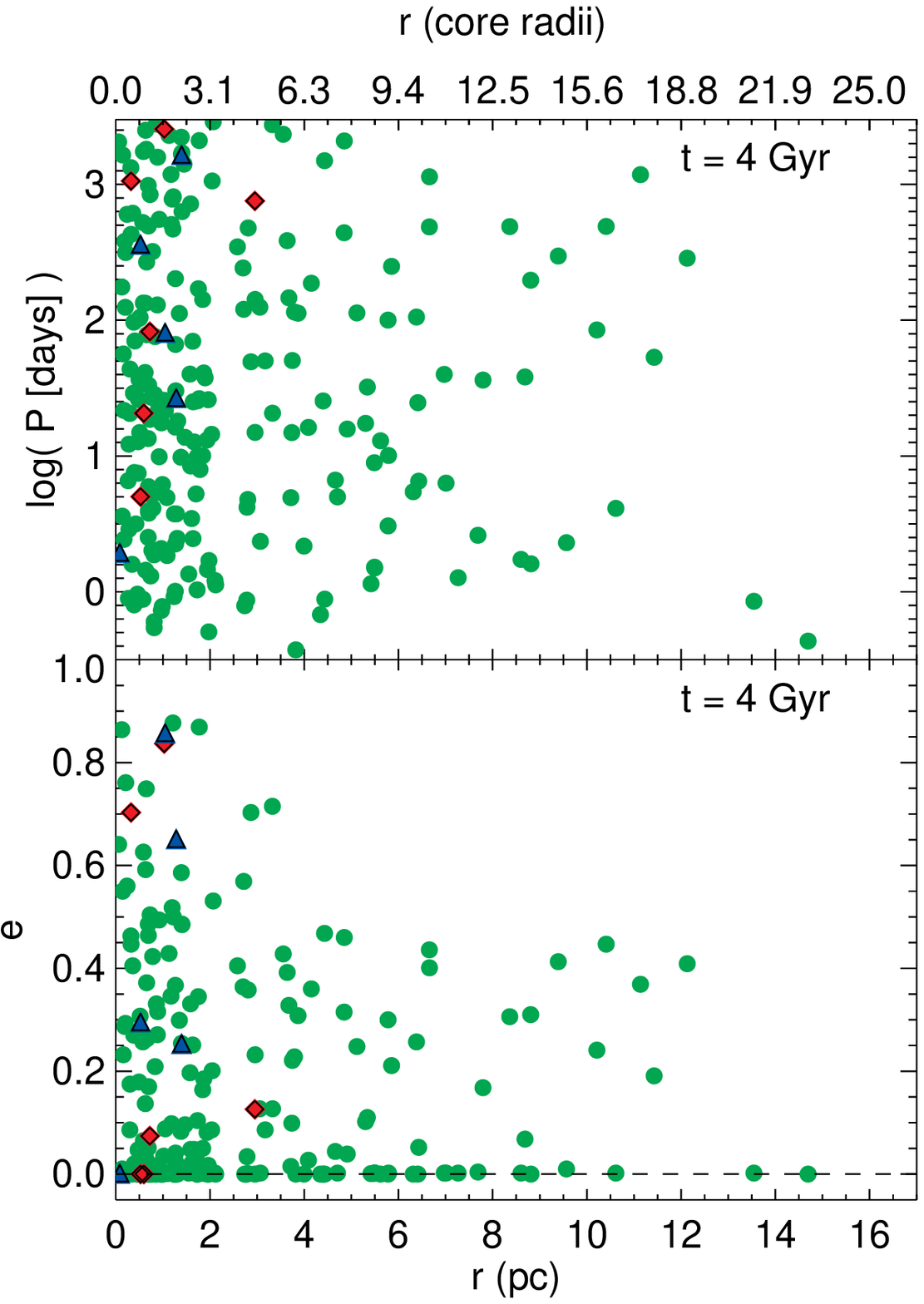}
\caption{\footnotesize Log period (top) and eccentricity (bottom) plotted against projected radius from the cluster center for 
the solar-type initial (left) and final (at 4 Gyr; right) binary populations.
We show the main-sequence, giant and blue straggler binaries in light-gray circles, dark-gray diamonds and black triangles 
(green, red and blue in the online version), respectively.
(In the initial plots all binaries are on the main-sequence.)
On the top x-axes we convert parsecs to core radii using a core radius of 0.64 pc \citep{hur05}.
Despite becoming more centrally concentrated, the overall form of the period distribution does not change over 4 Gyr of evolution.
On the other hand, we notice a population of high-eccentricity binaries in the core of the cluster at 4 Gyr that were not present 
in the initial population. Their higher eccentricities are the result of dynamical encounters.}
\label{Hpevr}
\epsscale{1.0}
\end{center}
\end{figure*}

The MS period distribution of the simulation is significantly different from that observed in NGC 188 (Figure~\ref{Pfreq}).
In particular, the simulation contains a large overabundance of short-period MS binaries as compared to both NGC 188 and the 
Galactic field.
If we normalize the MS period distribution from the simulation to contain the same number of binaries as in the NGC 188 MS sample, 
a $\chi^2$ test shows that the two distributions are distinct at the $>$99\% confidence level.

The distribution of binary periods has not changed significantly 
over the 4 Gyr of simulated evolution.
A K-S test relating the initial and final hard-binary period distributions (Figure~\ref{HPfreq}) shows no distinction.
Thus this discrepancy between the simulated and observed populations is a result of the flat initial distribution 
in log($P$) used in the simulation.

\subsubsection{Orbital Eccentricity}

In Figure~\ref{HelogP} we show the $e - \log(P)$ diagram for the \citet{hur05} simulation.
%(again using the same gray scale coding for the  simulated MS, giant and BS populations as we have done throughout the paper for NGC 188.)
The $e - \log(P)$ distribution of the simulation MS binaries is strikingly different from the observed
$e - \log(P)$ distribution of MS binaries in NGC 188 (top panel, Figure~\ref{elogP}), and indeed from
observations of many similarly evolved open clusters and the Galactic field \citep[see][R10]{mei05}. The differences 
in the simulation include:

\begin{enumerate}
\item The upper envelope on orbital eccentricity, rising with increasing period. This derives directly from the thermal initial conditions and the Kroupa tidal evolution prescription. The observed distribution of orbital eccentricities longer than the circularization period is better matched with a wide Gaussian in eccentricity at all periods, as found in the young cluster M35 \citep{mat08} and in the field (R10), and as used in the modeling of detached binary evolution by \citet{mei05}.

\item The large number of circular binaries with periods greater than the observed circularization period
of 15.0 days. Many of these systems had circular orbits initially due to the Kroupa pre-evolution process. An additional small population of 
long-period circular binaries was created through unstable mass transfer (common-envelope) events involving initially non-circular MS
binaries. 

\item The large number of eccentric binaries with periods shorter than the circularization period.  This suggests 
that the tidal energy dissipation rate is underestimated in the \citet{hur05} model.  Indeed \citet{bel08} suggest 
that the convective tidal damping should be increased by a factor of 50 to match observations of M67.
\end{enumerate}

Similarly, in Figure~\ref{Hefreq} we show the eccentricity distributions from the simulation. 
%We use the same coding as in Figure~\ref{HPfreq}. 
Here we use the same period limits of $P_{circ} < P < 3000$ days as in 
Section~\ref{Sefreq} to define our sample, where $P_{circ}$ = 15.0 days (for NGC 188 found by \citealt{mat04}).
The marked excess in the number of circular orbits compared to NGC 188 (Figure~\ref{efreq}) and the field (R10) is evident. This same excess is seen in comparison with the young open cluster M35 \citep{mat08}. Most of these circular orbits are in short-period binaries, and as such play a critical role in the formation of BSs in the simulation.

\begin{figure*}[!t]
\begin{center}
%\plottwo{f16a.eps}{f16b.eps}
\plottwo{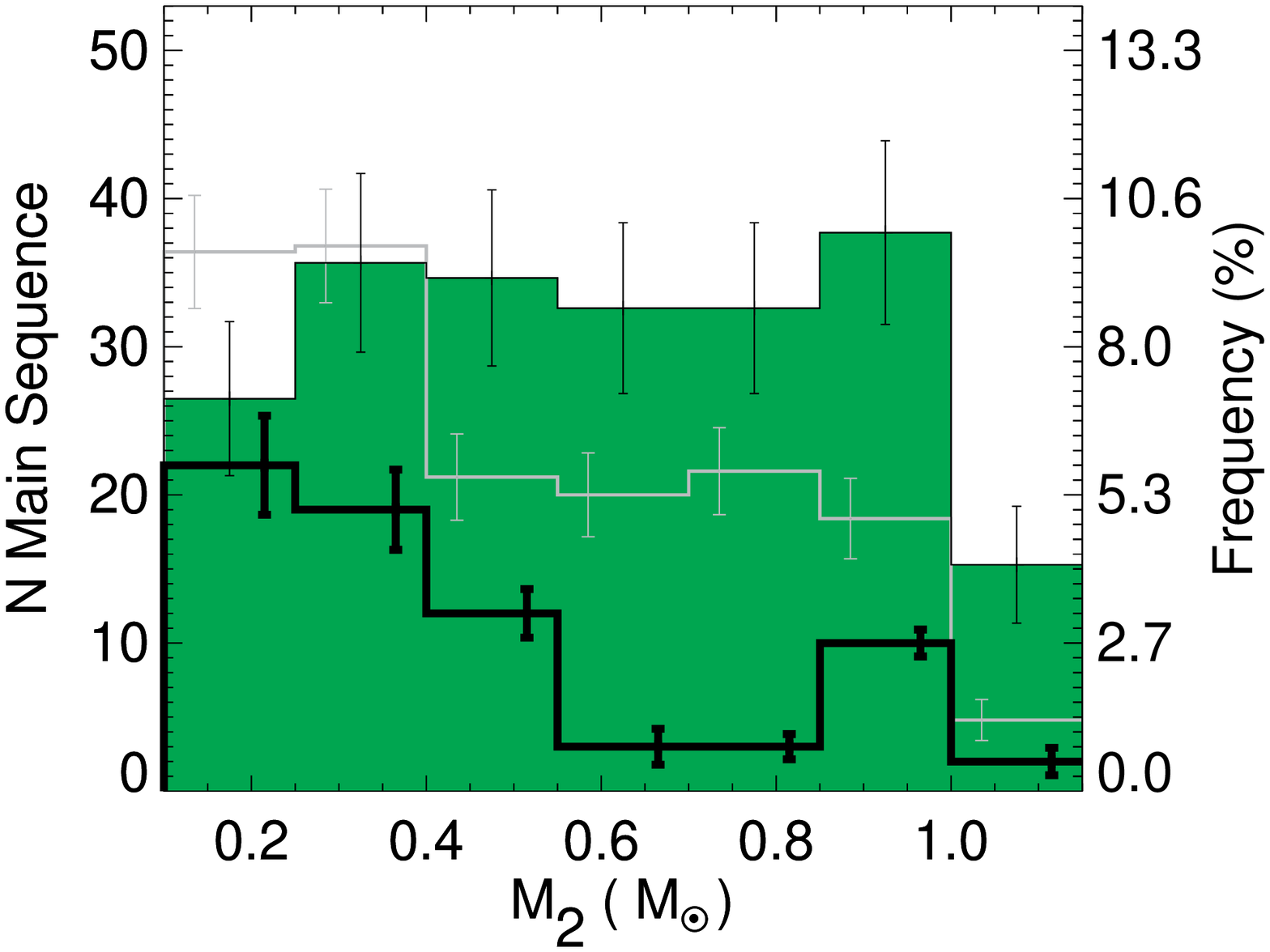}{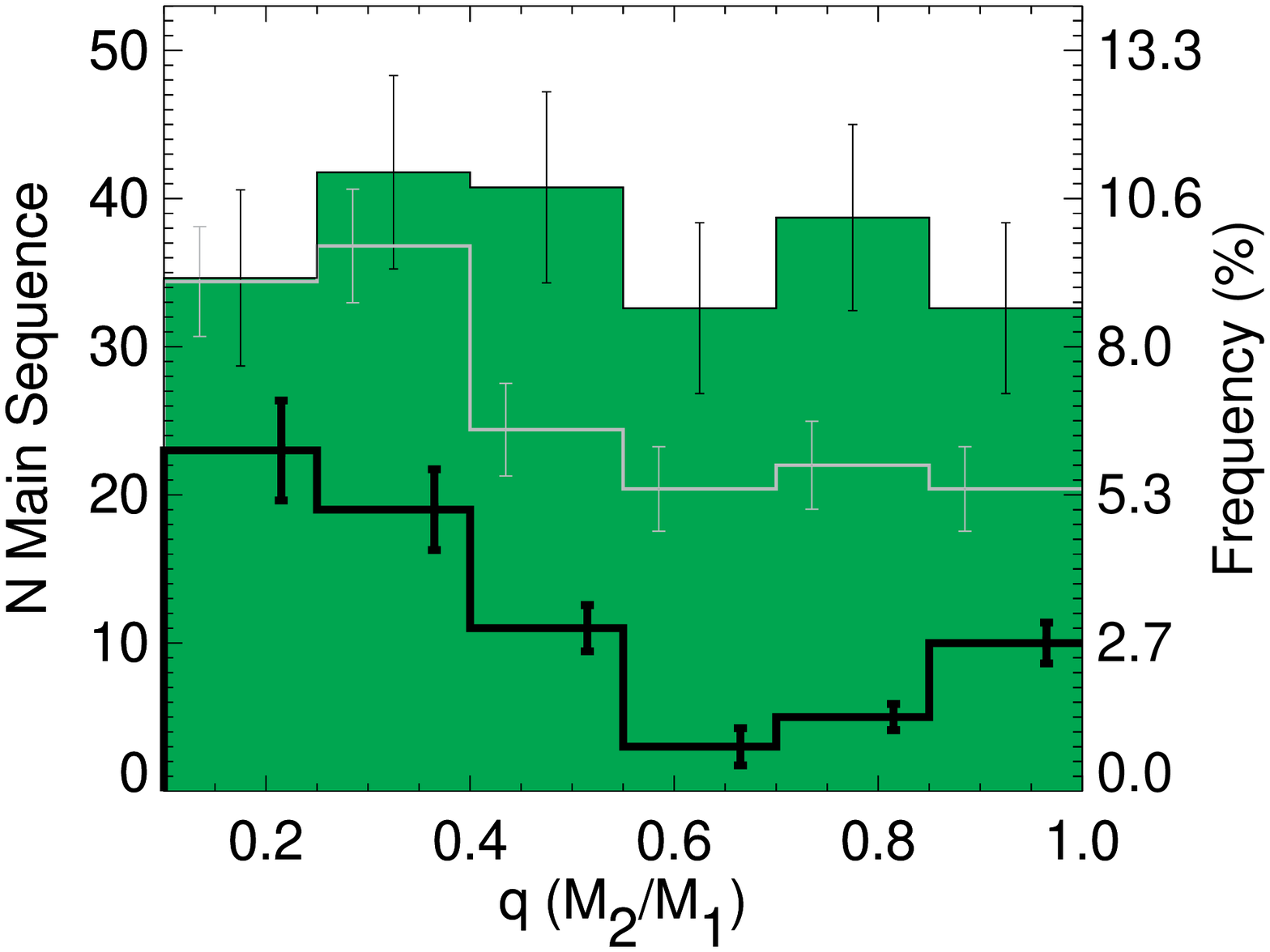}
\caption{\footnotesize Secondary-mass (left) and mass-ratio (right) distributions for solar-type main-sequence
binaries in the \citet{hur05} simulation (filled histograms) and in NGC 188 (thick-lined black histograms), both for binaries 
with $P < 3000$ days.
We also show the initial populations from the model within the same mass and period range in the gray lines.
We normalize the simulated distributions by the NGC 188 sample size, and show the true frequencies on 
the right y-axes.
As the simulation only included stars with masses $>$0.1 \Msolar~in the initial population, 
we show the secondary-mass distribution from 0.10 - 1.15 \Msolar~and the mass-ratio distribution from 0.1 - 1.}
\label{Hqm2freq}
\epsscale{1.0}
\end{center}
\end{figure*}

Finally, in Figure~\ref{Hpevr} we show the initial and final distributions of eccentricity as functions of radius.
Over the course of the cluster evolution, a population of higher-eccentricity binaries developed 
within $\sim$5 core radii from the cluster center. (This population can also be seen in the $e - \log(P)$ diagram
of Figure~\ref{HelogP}.) No similarly high eccentricity binaries are found in the cluster halo. 
The increases in the orbital eccentricities of these binaries can be traced to dynamical encounters. 
These high-eccentricity binaries in the model are likely analogues of the high-eccentricity binaries observed 
in the core of NGC 188 (discussed in Section~\ref{Sefreq}), suggestive of a dynamical origin for some or 
all of the NGC 188 high-eccentricity binaries.

\subsubsection{Secondary-Mass and Mass-Ratio Distributions}

Unlike our observed NGC 188 sample where we require a statistical method to derive the secondary-mass and 
mass-ratio distributions, within the simulation we know the secondary masses and mass ratios for every binary.  
In Figure~\ref{Hqm2freq} we show the 4 Gyr secondary-mass (left) and mass-ratio (right) distributions for the simulated MS
binaries (filled histograms), and compare to the respective NGC 188 
MS distributions (thick-lined black histograms) as well as the initial distributions in the simulation (thin-lined gray histograms). The excess in hard-binary frequency - both initial and at 4 Gyr - in the simulation compared to NGC 188 is again clearly evident. However here we focus on the shapes of the distributions.

Both the secondary-mass and mass-ratio distributions of the simulation are essentially flat at 4 Gyr, except for the highest mass secondaries which are few in the simulation. This flat shape is in marked contrast to that found in NGC 188, which rises to lower masses and mass ratios. The NGC 188 distribution also shows some evidence of a peak at comparable primary and secondary masses ("twins"), which is not seen in the simulation.

In fact, the initial secondary-mass and mass-ratio distributions of the simulation do rise to lower secondary mass and look much like the current NGC 188 distributions. However, by 4 Gyr they flatten with subsequent dynamical evolution of the cluster. This would suggest that the NGC 188 initial distribution may itself have been even more peaked to lower masses than currently observed. However the simulation is silent with respect to the creation or evolution of a population of twins. 

The 4 Gyr mass-ratio distribution from the simulation is quite similar to that of the field binaries of R10, in that the distribution is consistent with being uniform over mass ratios of $\sim$0.2 - 0.95.  However the simulation does not reproduce the abundance of twins ($q \sim 1$), again suggesting that this feature seen in both the field and open clusters is likely linked to the processes of binary formation rather than dynamics.

\subsubsection{Summary}

The MS hard-binary population predicted by the simulation is significantly different from the MS hard-binary population of NGC 188. Along some dimensions this reflects on the adopted initial conditions of the simulation and along others it reflects on the physics of the simulation itself.

The initial binary frequency and period distribution of the simulation are largely maintained throughout the cluster lifetime, and so the marked differences with the NGC 188 frequency and period distribution are the result of inaccurate initial conditions. Specifically, the flat initial period distribution is distinct from the log normal period distribution of the observed young ($\sim$150 Myr) open cluster M35 \citep{mat08}, and the initial hard-binary frequency integrated over period is a factor two higher than found in such clusters. Given that the initial binary period distribution imprints itself on the cluster throughout its evolution, it is imperative that the properties of the initial binaries used in future $N$-body simulations be guided by observations of such young clusters.

The differences in the simulated and observed orbital eccentricity and $e - \log(P)$ distributions are also in large part the result of inaccurate initial conditions. Again, the eccentricity distributions in observed open clusters (e.g., M35, \citealt{mat08}; and see \citealt{mei05})
argue that a Gaussian-like distribution in eccentricity, with a weak dependence on period above the initial circularization cutoff, is a better choice than a thermal distribution.

In contrast, the large frequency of eccentric binaries with periods shorter than the circularization period of M67 and NGC 188 indicates that the physics of tidal circularization incorporated in the simulation is not accurate. Arguably the tidal energy dissipation rate is too small, as suggested by \citet{bel08}. This can be empirically tested with future simulations. On the other hand, the Kroupa (1995) tidal evolution prescription applied to the initial binary population evidently produced too many initial circular orbits at longer periods. 

The presence in the simulation of a central concentration of binaries and especially binaries in the core with dynamically produced high orbital eccentricities is in agreement with the binary population of NGC 188 and is encouraging with respect to the dynamical encounter physics in the simulation. We have not attempted a specific comparison of rates and frequencies, in part because \citet{hur05} derive a core radius for the simulation of 0.64 pc at 4 Gyr,
smaller than observed in NGC 188 (1.3 pc, \citealt{bon05}; 2.1 pc, \citealt{chu10}). This higher central concentration
will change evolution timescales in ways that likely are not easily compensated for accurately.

The correct initial secondary-mass and mass-ratio distributions remain poorly known, and comparison with NGC 188 provides only a rather distant lever on the question. Based on the simulation and NGC 188 distributions, it would appear that a secondary distribution rising to smaller masses is appropriate, and perhaps more so than used by Hurley et al. (2005). More intriguing is whether twins are present initially, given the hint of an excess of them existing both in NGC 188 and the field.

The CMD was the primary interface with observation used by Hurley et al. (2005), and they were very successful in matching the observed CMD of M67. In particular they reproduced well the BS population of M67. Importantly, the BS population of the simulation derived largely from mergers of close binaries, which we find to be over-represented in the simulation due to the adopted initial binary population. A plausible hypothesis is that reducing the binary frequency by a factor of two and reducing the frequency of the closest binaries relative to those of longer period would substantially reduce the BS production rate and under-predict the BS population compared to M67. Indeed, \citet{hur05} note that their $N$-body models using initial period distributions such as \citet{duq91} and \citet{kro95b} produce significantly fewer BSs than their M67 model (which used a flat initial period distribution).
Thus such models may require an offsetting change in the physics of BS formation or the relative frequency of BS formation channels in order to recover agreement with M67. We begin discussion of this issue in the next section.

\subsection{Comparison of the Blue Straggler Populations in NGC 188 and in the Simulation} \label{ShBS}

\subsubsection{Comparison of the Blue Stragglers in Binaries}

\begin{figure}[!t]
\begin{center}
\plotone{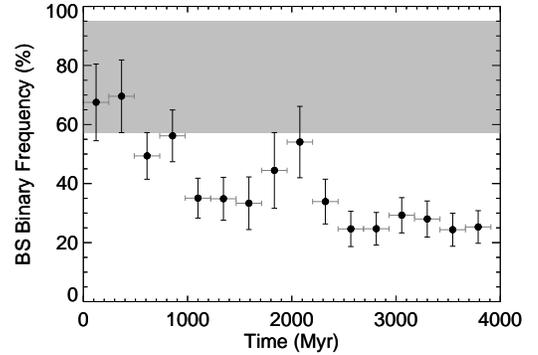}
\caption{\footnotesize Blue straggler hard-binary ($P < 10^4$ days) frequency as a function of time in the
\citet{hur05} $N$-body simulation.  To reduce the uncertainties and scatter we average every four adjacent 
time steps, so that each point represents $\sim$240 Myr in time.  We show the region in time covered by each point 
in the gray horizontal bars.  For comparison, we also show the NGC 188 BS binary frequency of 76~$\pm$~19~\% with the 
light gray band.}
\label{M67hBSfreq}
\end{center}
\end{figure}

Using the integrated sample of BSs, 
we derive a BS hard-binary ($P<10^4$ days) frequency of 29~$\pm$~2~\% (168/584). 
This hard-binary frequency for the BSs in the simulation is significantly lower than the 
NGC 188 BS hard-binary frequency of 76~$\pm$~19~\%.  
In Figure~\ref{M67hBSfreq}, we plot the \citet{hur05} BS hard-binary frequency as a function of cluster age
(no longer limited to ages $>$2 Gyr).
Over most of the lifetime of the simulated cluster the BS hard-binary frequency is less than half that of NGC 188 (and M67).
Thus, although the simulation has an overabundance of hard MS binaries, they are decidedly lacking amongst the simulated 
BS population. 
 
We estimate from the CMD location that the most massive NGC 188 BS in a binary with an orbital solution has a mass of $\sim$1.5 
times the MS turnoff mass of 1.1 \Msolar. 
The MS turnoff mass in the simulation is 1.7 \Msolar~at 2 Gyr and decreases to 1.3 \Msolar~by 4 Gyr.
Roughly 44\% of the BSs in binaries ($P < 3000$ days) in the simulation (between 2 - 4 Gyr) have masses $>$1.5 times 
the respective simulation turnoff mass.  Thus, on average the BSs in the simulation are more massive than those in NGC 188.

\begin{figure*}[!t]
\begin{center}
%\plottwo{f18a.eps}{f18b.eps}
\plottwo{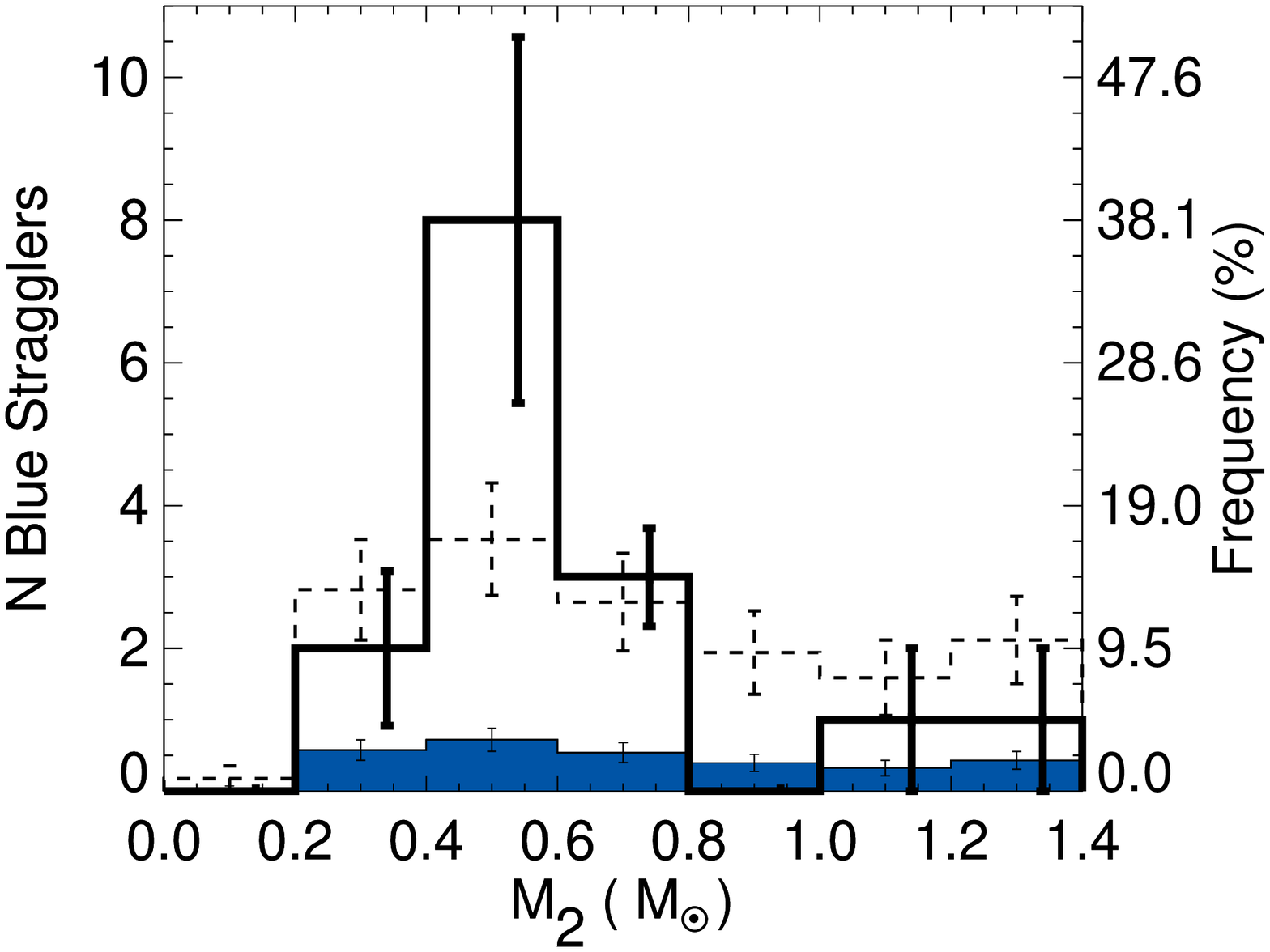}{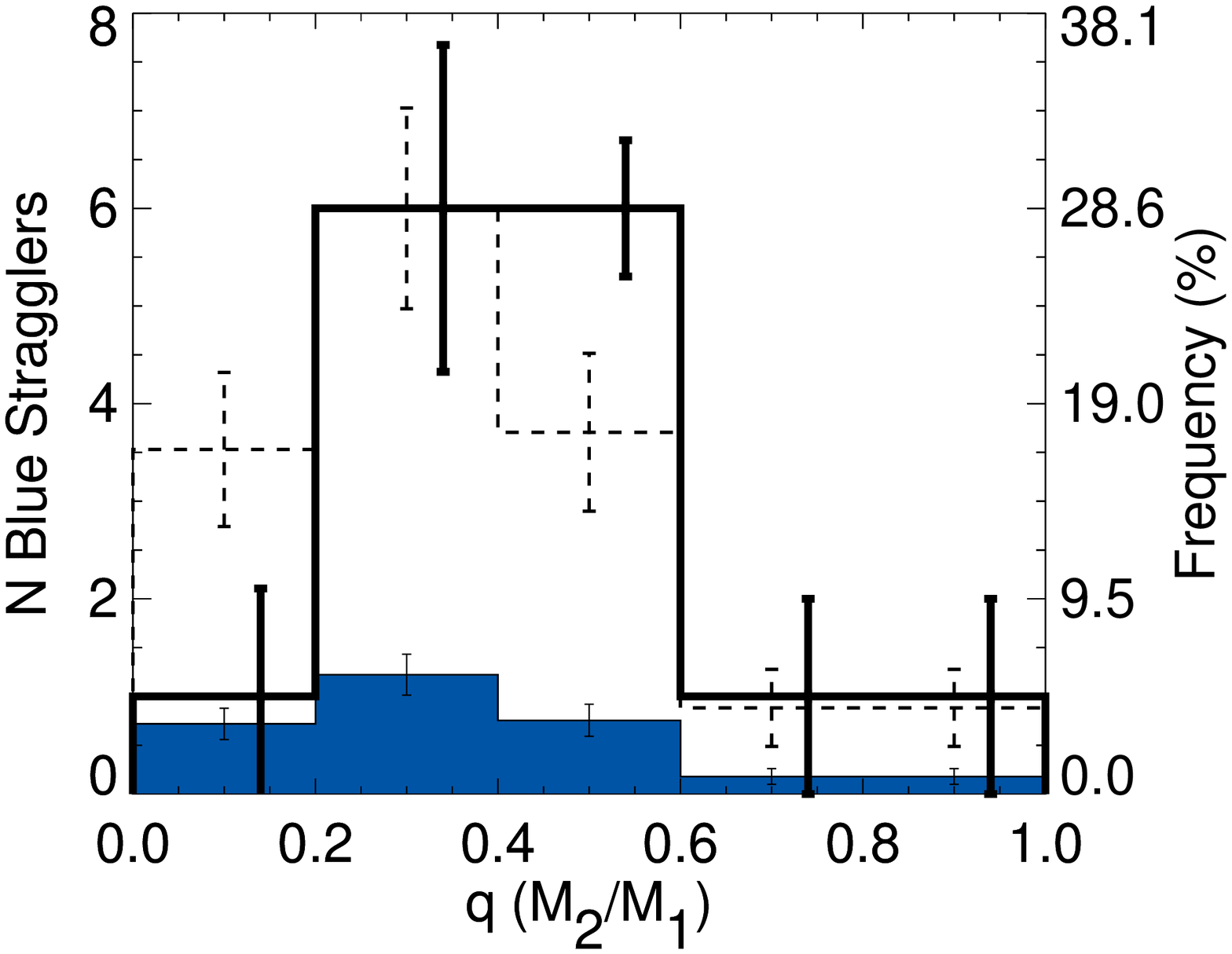}
\caption{\footnotesize Secondary-mass (left) and mass-ratio (right) distributions comparing the blue straggler binaries
in the \citet{hur05} simulation to those in NGC 188.
The thick-lined histograms reproduce the results shown in Figure~\ref{qm2freqBS} for the NGC 188 blue stragglers. 
For the simulation, we use the integrated sample of all simulated BSs from 2 - 4 Gyr.
The filled histograms show the simulated distributions normalize by the NGC 188 sample size.  
The true frequencies are shown on the right y-axes. For comparison of distribution shapes we also re-normalize the distributions from 
the simulation to contain the same number of blue straggler binaries as in NGC 188, shown with the dashed lines.}
\label{Hqm2freqBS}
\epsscale{1.0}
\end{center}
\end{figure*}

In Figure~\ref{Hqm2freqBS} we compare the NGC 188 BS secondary-mass and mass-ratio distributions to those in 
the \citet{hur05} simulation.  
The BSs in the simulation do not show the strong peak at $\sim 0.5$~\Msolar~that is observed in the NGC 188 BSs.
Instead the distribution is roughly uniform for secondary masses $>$0.2~\Msolar.
Interestingly the observed and simulated mass-ratio distributions for the BSs appear to be quite similar in form.
This agreement between the mass-ratio distributions is the result of dividing high-mass companions 
by the high-mass BSs, both of which individually are inconsistent with the NGC 188 BSs.

The binaries in the integrated BS sample of the simulation also have significantly different periods and eccentricities than the BS 
binaries of NGC 188. Specifically, the simulation produces BS binaries with shorter periods and a higher frequency of large eccentricities.
A 2D K-S test comparing the observed and simulated BS $e - \log(P)$ distributions 
shows that the two distributions are drawn from different parent populations at the $>$99\% confidence level.
Likewise 1D K-S tests comparing the observed and simulated eccentricity and period distributions return distinctions
at the 99\% and $>$99\% confidence levels, respectively.

Thus the BS sample from the simulation has significantly different binary properties
from those of the NGC 188 BS population.  These discrepancies are almost certainly linked, at least in part,
to the inaccurate initial binary population used in the simulation.

\subsubsection{Comparison of the Blue Straggler Spatial Distributions} \label{SBSspat}

\begin{figure}[!t]
\begin{center}
\plotone{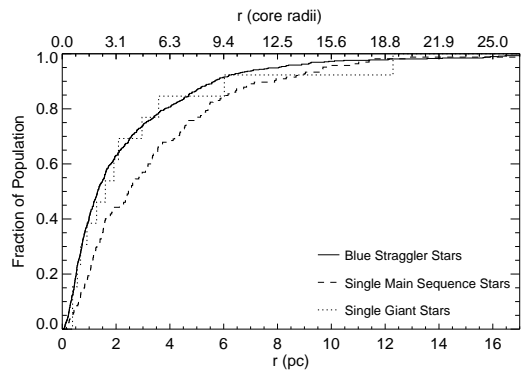}
\caption{\footnotesize Cumulative spatial distribution of stellar populations in the \citet{hur05} $N$-body model.
The blue stragglers from the integrated sample are shown in the solid line.  This BS spatial distribution is compared to the 
single main-sequence and giant stars, shown in the dashed and dotted lines, respectively.  The blue stragglers are 
more centrally concentrated than the single main-sequence stars (at the 98\% confidence level). }
\label{BSspatdist}
\end{center}
\end{figure}

The spatial distribution of the BSs in the simulation is also quite different from what we observe in NGC 188 (Figure~\ref{BSspatdist} 
and see also Figure 9 from Paper 1).
In Paper 1 we show that the NGC 188 BSs have a seemingly bimodal spatial distribution, with a population of 14 centrally concentrated 
BSs in the core and 7 BSs in the halo.  The full NGC 188 BS population is not significantly more centrally concentrated than the 
solar-type single stars in the cluster.
In the simulation, we do not observe this relatively large population of BSs towards the cluster 
halo.  Instead the full BS population is centrally concentrated with respect to the single MS stars (at the 98\% confidence level; also see Figure~\ref{Hpevr}).
Furthermore, if we only analyze the 4 Gyr BS population, all but one BS is within 4.5 pc from the cluster center; the remaining BS is
at a radius of 19.38 pc which is beyond the spatial extent of our NGC 188 survey.

Other than the difference in number of BSs, the core and halo BS populations in NGC 188 are statistically indistinguishable.  
Both have consistent binary frequencies 
and indistinguishable distributions of orbital eccentricities, periods and mass functions.  Furthermore both the core and halo 
populations contain one of the two short-period SB2 BSs, respectively.  The distributions of $v \sin i$ rotational velocities \citep{mat09}
are also consistent between the core and halo populations.

Similar bimodal BS spatial distributions have also been observed in globular clusters \citep[e.g.,][]{fer97,zag97,fer04,sab04,lan07}.
One interpretation for the origins of these peculiar spatial distributions has been the presence of two distinct formation mechanisms, with 
mass transfer dominating in the halos and collisions dominating in the core \citep{map06}.  Alternatively the distribution
may be a dynamical signature where encounters kick BSs out of the core leaving them to either relax back towards the center or 
remain in the halo depending on the local relaxation time \citep{sig94}.  

\begin{figure*}[!t]
\begin{center}
\plottwo{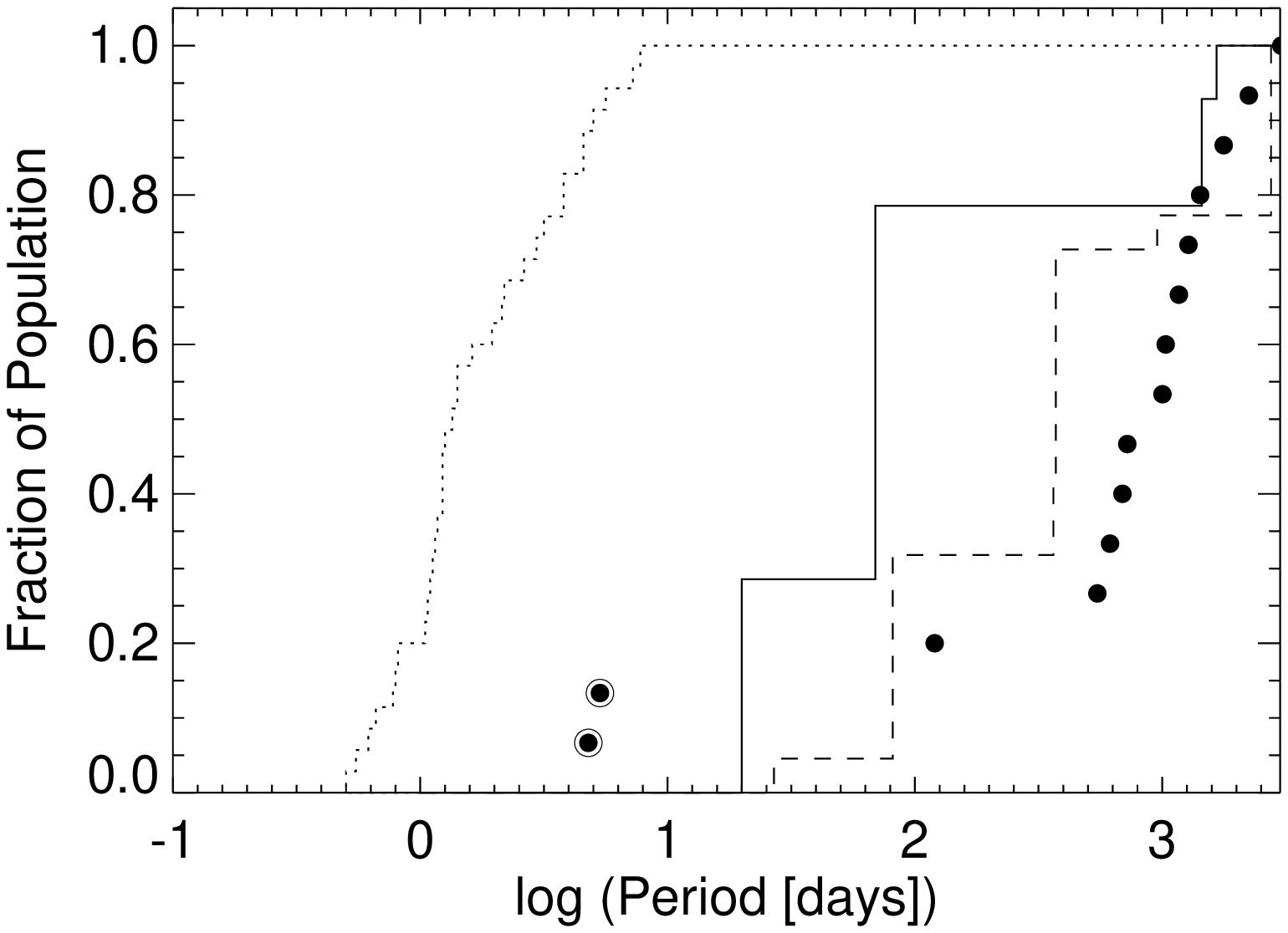}{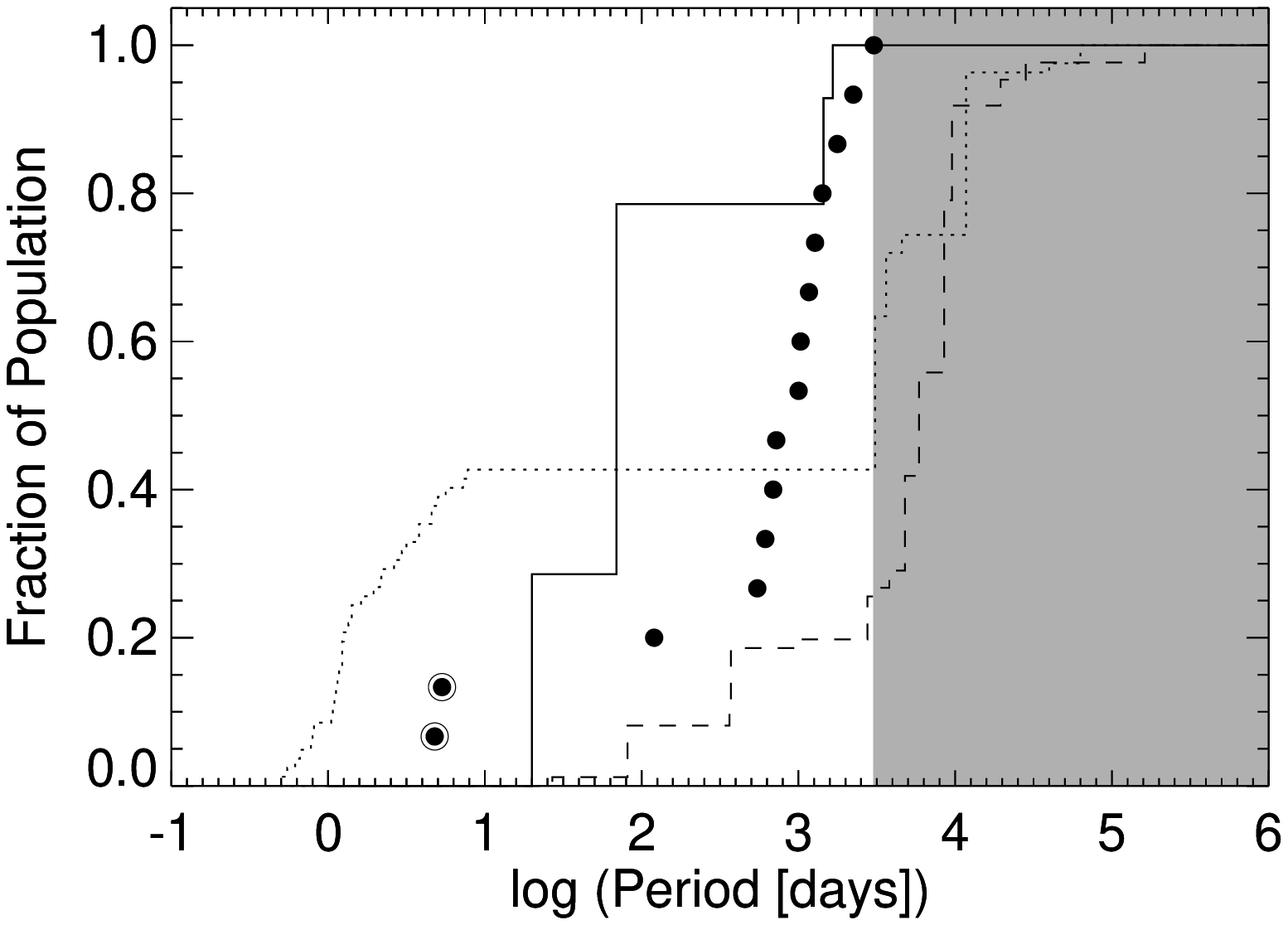}
\caption{\footnotesize Cumulative orbital period distributions comparing the NGC 188 blue stragglers to those in the \citet{hur05} simulation.
The NGC 188 blue straggler orbital periods are shown in the black points, with the SB2s circled.  The period distributions
from the simulation are divided by formation channel, with mass transfer in the solid line, mergers in the dotted line and collisions
in the dashed line. 
In the left panel we show only blue straggler binaries in the integrated sample from the model with $P < 3000$ days, while on 
the right we no longer limit by period.
In the right panel the gray-filled region marks periods beyond our observational completeness limit.
The simulation predicts that merger products will either be found in contact binaries (in the process of merging) or in 
wide binaries that result from exchange encounters, both of which are inconsistent with the NGC 188 blue straggler periods.
Collision products that remain in binaries are predicted to have a significantly lower frequency of binaries with $P<3000$ days
than observed for the NGC 188 blue stragglers.
The mass-transfer products, on the other hand, all have periods within our completeness limit.}
\label{BSPcum}
\end{center}
\end{figure*}

The similarities in the observed properties between the core and halo NGC 188 BS samples suggest that in NGC 188 this spatial 
distribution is not the result of two distinct formation mechanisms.  Instead, as discussed in \citet{gel11}, the majority of the 
NGC 188 BSs are consistent with an origin in mass transfer.
Dynamical encounters in the core similar to those suggested by 
\citet{sig94} (though not resulting in collisions) may be responsible for the BS spatial distribution.  
Alternatively the majority of NGC 188 BSs may have formed with minimal dynamical disturbance following a very similar 
spatial distribution to the binaries in the cluster.  The ``gap'' now seen in the spatial distribution 
between radii of $\sim$5 and $\sim$15 arcmin may then be a result of mass segregation processes, and may  
mark the region where the relaxation time approaches the age of the BSs.  
(A similar mechanism has also been suggested by other authors, e.g., \citealt{map04} and \citealt{lei11}, to explain the bimodal 
BS spatial distributions observed in some globular clusters.)

\subsubsection{Testing Blue Straggler Formation Channels in the Cluster Simulation} \label{BSformch}

There are three mechanisms for forming BSs in the \citet{hur05} $N$-body model: mergers, mass-transfer processes and 
stellar collisions.  Here we investigate the binary populations for BSs formed from each mechanism as they appear 
in the integrated sample, and compare their binary characteristics
to those observed for the NGC 188 BSs.  
In addition to completed mergers, we include in the merger category any binaries that are currently in contact or 
undergoing mass transfer and eventually merge. 
(In the \citealt{hur05} model, mergers result from Case A and/or B mass transfer.  The 
model never contains a large enough population of triple stars to test BS formation through the recent 
Kozai-based merger mechanism of \citealt{per09}.)
Between 2 and 4 Gyr in the simulation, 75\% (439/584) of the BSs in the integrated sample formed through mergers, 
20\% (117/584) of the BSs formed through collisions
and 5\% (28/584) of the BSs formed through mass transfer (that did not result in a merger).

The relative rates of BSs created through each of these processes in the simulation are closely linked to the initial binary population.
The dominant formation channel for the BSs in the simulation is mergers, a direct consequence
of the over-abundance of short-period binaries included in the primordial population.  We've shown that the initial binary 
population used in the simulation is not appropriate for NGC 188; thus the absolute and relative BS formation \textit{rates} cannot 
be expected to agree with the NGC 188 BSs.

However the predictions for the BS binary \textit{properties} of each formation channel (e.g. distributions of binary
orbital parameters) are only minimally sensitive to the initial binary population, and therefore
still provide a valuable means to investigate the origins of the NGC 188 BSs.
Here we use the \citet{hur05} models to define theoretical expectations for properties of the BSs formed
by each channel and compare to the observed properties of the NGC 188 BSs. Our goal is to use these
properties to begin to determine which mechanisms played significant roles in the formation of the NGC 188 BSs.

We begin first with the BS binary frequency. As expected, the merger mechanism predicts a low BS binary frequency ($P<10^4$ days) 
of only 13~$\pm$~2~\% (61/439),
much lower than the observed NGC 188 BS binary frequency. Furthermore, the majority of these binaries are contact systems, 
which we do not observe among the NGC 188 BSs.

The collision mechanism predicts a total BS binary frequency of 74~$\pm$~8~\% (86/117). However, the vast majority (74\%; 64/86) 
of the BSs in binaries produced by collisions have periods longer than 3000 days, our completeness limit for orbital solutions in NGC 188 
(see Section~\ref{incomp}).  67\% of the NGC 188 BSs have periods shorter than 3000 days, which contradicts the 
simulation result of only 19\% (22/117) binaries with periods of less than 3000 days from the collisional formation mechanism.

The mass-transfer mechanism predicts a 100\% BS binary frequency, all of which are detached and have periods
of less than 3000 days.

Next we consider the distributions of orbital parameters predicted by the simulation for binaries formed by each mechanism, beginning
with the period distributions. The predicted cumulative period distributions for the three mechanism are shown in Figure~\ref{BSPcum};
the left panel is limited to BS binaries with periods below our completeness limit while the right panel shows
all BS binaries from the simulation.  

Two populations of merger products are 
evident in Figure~\ref{BSPcum}: a grouping of contact systems at short periods and a grouping at longer periods where the
BSs were subsequently exchanged into binaries.
The period distributions of both of these merger populations are inconsistent with the period distribution of the NGC 188 BSs. 

The distribution of collisionally produced BS binaries with periods of less than 3000 days is consistent with the shape of the
observed distribution of NGC 188 BS binaries. However, as already noted, the frequency of such binaries in the simulation is 
much too small.

\begin{figure}[!t]
\begin{center}
\plotone{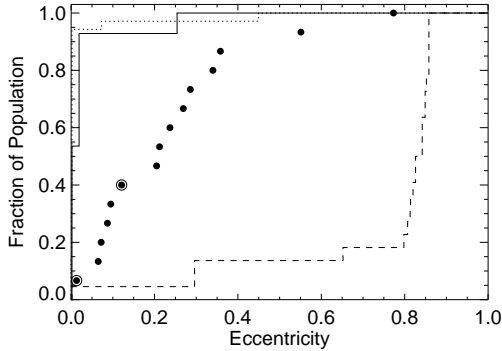}
\caption{\footnotesize Cumulative orbital eccentricity distributions comparing the NGC 188 blue stragglers to those in the \citet{hur05} simulation, for binaries with $P < 3000$ days.
The NGC 188 blue straggler orbital eccentricities are shown in the black points, with the SB2s circled.  The eccentricity distributions
from the simulation are divided by formation channel, with mass transfer in the solid line, mergers in the dotted line and collisions
in the dashed line. 
The $N$-body model predicts that collision products will have significantly higher eccentricities than observed in the NGC 188 
blue straggler binaries.  Both mass-transfer and merger products are predicted to have nearly all circular orbits.  
(The few non-circular orbits in these samples are the result of dynamical encounters.)  
The assumption of circular orbits for contact binaries is likely valid.
However the assumptions used in the \citet{hur05} model of rapid 
circularization during mass transfer may not be valid in all cases \citep{sok00,bon08,sep09}.  Thus the accuracy of the 
predicted eccentricity distribution for mass-transfer products from the model is uncertain.}
\label{BSecum}
\end{center}
\end{figure}

All of the mass-transfer BSs in the \citet{hur05} model 
have periods within our detection limit, and are split between one population near 100 days (a result of Case B mass transfer from 
a red giant) and a second population near 1000 days (a result of Case C mass transfer from an asymptotic giant).\footnote{\footnotesize 
Though Figure~\ref{BSPcum} appears to suggest that the \citet{hur05} model favors Case B over Case C mass transfer, the difference in 
frequency is instead due to two longer lived Case B products rather than a significantly larger number of BSs produced through Case B 
mass transfer.} 
Similarly all but one of the NGC 188 SB1 BSs are found with periods near 1000 days; the remaining SB1 BS has a period of $\sim$120 days.  
The theoretical and observed distributions are not formally distinguishable. 

The NGC 188 SB2 BS binaries both have very short periods around 5 days.  
Interestingly, the only formation mechanism in the M67 model that produces BSs with such short periods is the merger mechanism. 
However both NGC 188 BSs are in detached binaries.  Thus the M67 model struggles to produce BSs with similar characteristics
to the two SB2 BS binaries in NGC 188.

\begin{figure}[!t]
\begin{center}
\plotone{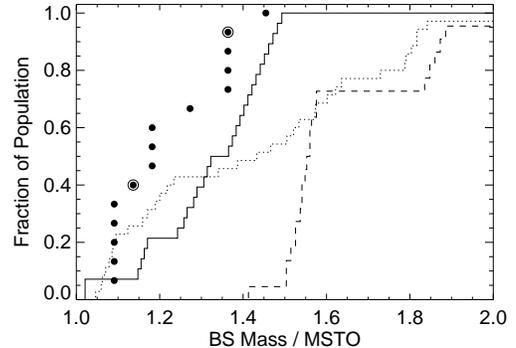}
\caption{\footnotesize Cumulative blue straggler mass distribution comparing the NGC 188 blue stragglers to those in the \citet{hur05} simulation.
We normalize all masses by the respective main-sequence turnoff mass to facilitate the comparison.
The integrated sample of blue stragglers from the simulation is limited here to only include binaries with $P < 3000$ days to 
reflect our observational completeness in orbital solutions.
Our NGC 188 blue straggler mass estimates are shown in the black points, with the SB2s circled.  The primary mass distributions
from the simulation are divided by formation channel, with mass transfer in the solid line, mergers in the dotted line and collisions
in the dashed line.  The simulation predicts that collision products are the most massive, followed by mergers and then mass-transfer
products.  
Both the collision and merger products are significantly more massive than the NGC 188 blue stragglers, while products of mass-transfer 
have a mass distribution that is statistically indistinguishable from that of the NGC 188 blue stragglers.}
\label{BSM1cum}
\end{center}
\end{figure}

The observed distribution of orbital eccentricities is a challenge for all of the formation mechanisms.
Collision products that retain a binary companion are predicted to have significantly higher eccentricities than 
observed for the NGC 188 BSs; the simulation result is rejected at the $>$99\% confidence level (Figure~\ref{BSecum}).  
Conversely, the mass-transfer and merger mechanisms both predict that essentially all of the BSs in binaries should have circular orbits.  
In the model, any non-zero orbital eccentricity prior to the BS formation is quickly circularized by tides during the mass-transfer 
or contact binary stages. The few non-circular orbits in both of these samples are the results of subsequent dynamical encounters.

Three of the SB1 BSs in NGC 188 do have orbits consistent with being circular. Theoretical main-sequence tidal circularization timescales 
for binaries at $\sim$1000-day periods are prohibitively long and cannot explain these circular orbits \citep{mei05}. Indeed, there are no circular MS binaries with periods near 1000 days in NGC 188. Thus these BS circular orbits may be evidence of a mass transfer origin.

However, the other 10 long-period BS binaries in NGC 188 have measurable eccentricities, one as high as 0.8. 
Recently, observational and theoretical evidence has begun to suggest that mass transfer, particularly 
involving an asymptotic-giant star, will not always lead to rapid circularization. 
Examples of eccentric post-mass-transfer systems may include barium-star systems \citep{jor98},
post-AGB binaries \citep[][and references therein]{van99}, and possibly also the long-period eccentric BSs 
in the galactic halo \citep{car05}. Theoretical work has suggested mechanisms that can yield ``eccentricity-pumping''
\citep[e.g.][]{sok00,bon08,sep09}, which were not included in the \citet{hur05} model. Thus at this time the implications of the observed
eccentricity distribution for the mass-transfer channel remains uncertain.

The eccentricities of the two short-period SB2's are interesting. The double BS binary 7782 found in the cluster halo has a circular orbit,
while the BS-MS pair 5078 has a small eccentricity of 0.17. Both are discussed in more detail in \citet{mat09}.
7782 is likely the result of a dynamical encounter and exchange that may have ejected it to the halo; presumably after such
an encounter it would have at least initially had a highly eccentric orbit.
On the other hand, 5078 has managed to sustain a small eccentricity against a theoretically short tidal
circularization timescale.

In the previous section we found that the BSs in binaries within the simulation are generally more massive than those in NGC 188, where 
all are $\lesssim$1.5 times the turnoff mass.  If we divide the integrated BS binary sample from the simulation by formation mechanism, we 
find that the mean masses from mergers, collisions and mass transfer are $1.43 \pm 0.05$, $1.72 \pm 0.10$ and $1.3 \pm 0.03$ 
times the turnoff mass, respectively.  Collision products in binaries are predicted to be the most massive BSs, followed by merger 
products and finally mass-transfer products.  

In Figure~\ref{BSM1cum} we compare the mass distributions from the simulated BSs to that of the NGC 188 BS binaries (with orbital 
solutions).
K-S tests show that the NGC 188 BSs are significantly less massive than predicted by the collisional and merger hypotheses, at the $>$99\%
and 95\% confidence levels, respectively.  
This finding is consistent with that of \citet{gle08}, who rule out recent formation of massive blue stragglers 
through collisions in NGC 188.
Thus both collisions and mergers generally produce more massive BSs 
than those observed in NGC 188.
On the other hand, a similar comparison between the predicted BS masses for mass-transfer 
products to the NGC 188 BS masses shows no significant distinction.  

Finally, and for completeness, we note that \citet{gel11} present an analysis of the
secondary-mass distributions of the SB1 BS binaries of NGC 188 (e.g., Figure~\ref{qm2freqBS}) and those formed in a new $N$-body simulation of 
NGC 188, created using a very similar code to that used by \citet{hur05}, but with a more realistic initial binary population. 
They find that the observed 
secondary-mass distribution, narrowly peaked around a mean mass of 0.53 \Msolar, 
is not consistent at $>$99\% confidence with the theoretical secondary-mass distribution of BS binaries 
formed by collisions, which have a mean mass nearly twice that observed. On the other hand, the observed secondary-mass 
distribution is fully consistent with a population of carbon-oxygen WDs with masses between 0.5~\Msolar~and
0.6 \Msolar. They also find that formation of BSs in hierarchical triples via the Kozai-induced merger mechanism is not favored
by the observed distribution, but is only rejected at the 98\% confidence level.

\subsubsection{Summary}

To summarize, our comparisons show that BSs formed by collisions that retain a companion will have a significantly lower
binary frequency ($P < 3000$ days),
longer orbital periods, higher orbital eccentricities, higher masses and higher-mass companions than are observed for the 
NGC 188 SB1 BS binaries. BSs formed by mergers as modeled by \citet{hur05} are also highly inconsistent with the
observations, most notably having a significantly lower binary frequency, shorter periods, and higher masses than are observed. 
Thus we can confidently
rule out both mergers within isolated close binaries and collisions as the dominant formation channel for the NGC 188 BS binaries.
Predictions for mass-transfer products are consistent with most of the binary properties of the NGC 188 SB1 BSs,
which comprise two-thirds of the NGC 188 BSs.

We are unable to investigate the Kozai-induced merger mechanism within the \citet{hur05} model, as this model never contains sufficient triples.
Importantly the Kozai-induced merger and mass transfer hypotheses make additional observable
predictions that can distinguish between the two mechanisms.  
Mass transfer predicts WD companions while the Kozai-induced merger mechanism predicts predominantly MS companions.
Relatively young ($\lesssim 0.4$ Gyr) WD companions are hot enough to contribute a significant excess in UV emission
to the combined spectrum of the BS binary, and can be detected through HST observations.
Additionally mass-transfer products are expected to show a higher depletion of lithium than merger and collision products
\citep{lom95,san97,nor97,gle10}.  (The field BS binaries for which lithium abundances are available all show
lithium depletion; \citealt{car01}.)  Mass-transfer processes may also produce modified CNO abundances \citep{she00}
and enhancements in $s$-process elements \citep{car01} from the processed material of the giant star donor.
Such observations of the NGC 188 BS population are currently underway and will soon help to further constrain the origins of the NGC 188 BSs.

However mass transfer cannot explain the companion masses or orbital periods of the NGC 188 SB2 BSs.
These two BS binaries along with S0182 of M67 \citep{san03} represent 
intriguing case-studies for future $N$-body simulations. We suspect that they are among the most direct 
observational evidence for the close dynamical encounters of binaries that $N$-body studies have predicted 
for decades.

Additionally there are five BSs in NGC 188 that aren't detected as binaries.  The M67 model suggests that 
these BSs likely did not all form through mass transfer, while both collisions yielding very wide companions and
mergers predict such BSs.

Finally, we anticipate that shifting the MS binary period distribution away
from short-period binaries (to better match the observations of MS binaries) will shift the dominant formation mechanism away 
from mergers and thereby change the properties of the resultant BS population, perhaps in better agreement with the binary observations. 
But recall that the \citet{hur05} simulation successfully matched the number and 
CMD distribution of the M67 BSs.  To retain this agreement in BS frequency with a reduced number of initial short-period binaries
may require modification of the physics and efficacy of the formation channels. 
We will explore this in a future paper discussing a more accurate $N$-body simulation of NGC 188.

\section{Summary and Conclusions} \label{summary}

Through a series of papers (Papers 1,2, \citealt{mat09} and \citealt{gel11}) of which this is 
the most recent contribution, we study the dynamical state of the old (7 Gyr) open cluster NGC 188.
In previous papers we identify a complete sample of single and binary cluster members
among the solar-type stars in NGC 188, present orbital solutions for the majority of our detected binaries, 
and study the BSs in detail.  Here we analyze the hard-binary population of the cluster, 
including the hard-binary frequencies and distributions of binary orbital parameters of the 
MS, giant and BS populations.  We compare these results in detail to the observed Galactic field binary 
population of R10 as well as the simulated binary population in the \citet{hur05} $N$-body open cluster 
simulation to study the impact that relaxation processes and close dynamical encounters may have on the 
cluster's binary population.

We find the properties of the solar-type MS binaries in NGC 188 to be consistent with similar 
binaries in the Galactic field.  
The NGC 188 MS binaries have a log-period distribution that increases towards our detection limit and 
is consistent with the log-normal distribution found by R10 for solar-type binaries in the Galactic field.
The MS eccentricity distribution has a roughly Gaussian form (for $P_{circ} < P < 3000$ days), also consistent with R10.
Likewise the MS binary secondary-mass and mass-ratio distributions are indistinguishable from those of similar binaries in 
the Galactic field.  In NGC 188 both the secondary-mass and mass-ratio distributions appear to rise towards lower masses 
and mass ratios, as also observed by \citet{maz03} and \citet{duq91} in the field.
Both NGC 188 distributions also show evidence for a secondary peak at mass ratios near unity, as observed by R10, \citet{tok00} 
and \citet{fis05} in the field.
The observed mass-function distribution for the NGC 188 MS binaries is formally consistent with a secondary-mass
distribution where secondaries are drawn from either a standard \citet{kro01a} IMF or a uniform mass-ratio distribution.

The observed global MS hard-binary frequency in NGC 188 is 23~$\pm$~2~\%, which when corrected 
for our incompleteness results in a true MS hard-binary frequency of 29~$\pm$~3~\%
for $P<10^4$ days.  This hard-binary frequency is marginally higher than that of the R10 field within 
the same period range, of 19~$\pm$~2~\%.  This somewhat higher hard-binary frequency in NGC 188 may 
be the result of dynamical relaxation processes where lower total mass single stars are preferentially 
ejected from the cluster, as is seen in $N$-body models such as \citet{hur05}.  
Two-body relaxation is evident in the central concentration of binaries in the cluster.  

Additional evidence for dynamical encounters may be seen in the population of high-eccentricity
binaries in the core of the cluster but not in the halo (Figure~\ref{evr}).  
High-eccentricity binaries such as these are created within the \citet{hur05} model in the cluster core
via dynamical interactions throughout the cluster lifetime.
  
Importantly, the NGC 188 BS binaries are significantly different from the NGC 188 MS 
solar-type binaries, providing some of the most distinguishing characteristics of the BS population found to date.
The observed BS hard-binary frequency in NGC 188 is 76~$\pm$~19~\%,  three times the observed solar-type 
MS hard-binary frequency.  The BS and MS hard binary frequencies are distinguishable at the $>$99\% confidence level.

The BS binary $e - \log(P)$ distribution is distinct from that of the MS at the 99\% confidence level.
The BS SB1 binaries in NGC 188 have particularly remarkable orbital characteristics.  All but one have periods 
near 1000 days, most have modest eccentricities (with three in circular orbits), and their secondary-mass distribution 
is narrow and peaked, with a mean mass of $\sim$0.5~\Msolar~(Figure~\ref{qm2freqBS}).  

The two short-period ($P<10$ days) BS binaries are both observed as SB2s.  
These SB2 BSs have the most massive companions to any of the BSs in NGC 188.
Indeed BS ID 7782 has a mass ratio of unity and is composed of \textit{two} BSs. 
We argue that these short-period BS binaries likely have a dynamical origin \citep{mat09}.

SB2 BS, ID 5078, has a companion star on the upper MS.
Normal stellar evolution tracks \citep{mar08} predict a mass for the BS primary star that is 
15\% to 30\% lower than the mass we derive using the kinematic mass ratio.
Thus normal stellar evolution tracks may underestimate the masses of BSs.

Comparisons to the BS population in the \citet{hur05} $N$-body open cluster model show that we can rule out 
the hypothesis that the NGC 188 SB1 BS binaries have a collisional origin.  Collision products that retain 
binary companions are predicted to have significantly higher eccentricities, higher masses,
higher-mass companions, and a lower frequency of binaries with $P < 3000$ days than observed for the NGC 188 SB1 BS binaries.

Likewise mergers within isolated binaries will not reproduce the binary properties of the NGC 188 BSs. 
BSs produced by isolated mergers are too massive, have a binary frequency that is much too low, and the few binaries have 
too short of periods.

Predictions from the mass transfer mechanism are consistent with most of the binary properties of the NGC 188 SB1 BSs, 
which comprise two-thirds of the NGC 188 BSs.  Mass transfer products are predicted to have a very high binary 
frequency, with long orbital periods, and companions of about 0.5~\Msolar, all closely consistent with the NGC 188 BSs 
(as also discussed in \citealt{gel11}).

The \citet{hur05} model never contains a sufficient frequency of triples to investigate the Kozai-induced merger mechanism for BS formation 
\citep{per09}. \citet{gel11} find that the observations do not favor this formation mechanism, but the data only reject the hypothesis at the 
98\% confidence level.  Spectroscopic observations to determine the abundances of the SB1 BSs as well as
HST observations aimed at detecting the FUV flux from the WDs predicted by the mass transfer mechanism are currently underway, and will 
further constrain the origins of the NGC 188 BSs.

Comparing detailed observations of open cluster binaries with predictions from sophisticated $N$-body open cluster models is a powerful method 
for investigating the dynamical evolution of a binary population and the formation of the BSs.
Comparisons with observations, like those presented in this paper, are also essential to determine the validity of the models.
In this paper we show that accurately defining the initial binary population is critical. We suggest
that this is best accomplished through comparisons with detailed observations of binary populations in 
young open clusters, although every indication to date is that the field binary population is a reasonable
proxy.
$N$-body simulations modeling NGC 188 and using an observationally defined initial binary population are underway and will 
soon allow for a detailed study of the dynamical evolution of the old open cluster NGC 188.

\acknowledgments
The authors express their gratitude to the staff of the WIYN Observatory without whom we would
not have been able to acquire the thousands of superb stellar spectra upon which this work rests. 
We also thank the many undergraduate
and graduate students who have helped to obtain these spectra over the years at WIYN for this project.
We owe many thanks to J.~Hurley for access to the output of his simulations and for many insightful conversations.
Thanks to T.~Mazeh for the helpful discussion on the mass-ratio and secondary-mass distributions.
This work was funded by the National Science Foundation grant AST-0908082, the Wisconsin Space Grant Consortium, and 
the Lindheimer Fellowship at Northwestern University.

Facilities: \facility{WIYN}, \facility{DAO:1.22m}, \facility{Hale}

\bibliographystyle{apj}                       %% AASTeX
\bibliography{ms.v3}

\end{document}